\documentclass{svjour3}                    
\smartqed  
\usepackage{natbib}
\usepackage{graphicx}
\usepackage{mathptmx}      
\usepackage{amsmath}
\usepackage{amssymb}
 \journalname{Celestial Mechanics \& Dynamical Astronomy}
\begin{document}

\title{Resonances of low orders in the planetary system of HD37124}

\titlerunning{Resonances in the system of HD37124}

\author{Roman V. Baluev}

\authorrunning{R.V. Baluev}

\institute{R.V. Baluev \at
           Sobolev Astronomical Institute, Saint Petersburg State University,\\
           Universitetskij prospekt 28, Petrodvorets, Saint Petersburg 198504, Russia\\
           \email{roman@astro.spbu.ru}
}

\date{Received: April 8, 2008 / Revised: June 21, 2008 / Revised: September 1, 2008 / Accepted: September 10, 2008}

\maketitle

\begin{abstract}
The full set of published radial velocity data ($52$ measurements from Keck + $58$ ones
from ELODIE + $17$ ones from CORALIE) for the star HD37124 is analysed. Two families of
dynamically stable high-eccentricity orbital solutions for the planetary system are found.
In the first one, the outer planets c and d are trapped in the 2/1 mean-motion resonance.
The second family of solutions corresponds to the 5/2 mean-motion resonance between these
planets. In both families, the planets are locked in (or close to) an apsidal corotation
resonance. In the case of the 2/1 MMR, it is an asymmetric apsidal corotation (with the
difference between the longitudes of periastra $\Delta\omega\sim 60^\circ$), whereas in
the case of the 5/2 MMR it is a symmetric antialigned one ($\Delta\omega = 180^\circ$).

It remains also possible that the two outer planets are not trapped in an orbital
resonance. Then their orbital eccentricities should be relatively small (less than, say,
$0.15$) and the ratio of their orbital periods is unlikely to exceed $2.3-2.5$.
\end{abstract}

\section{Introduction}
\label{intro}
For now, the star HD37124 is believed to host three Jovian planets. The innermost planet
`b' was discovered by \citet{Vogt00}. This planet moves on a low-eccentric orbit with a
period of $P_b\approx 150$~days. Soon after this, the second planet `c' was discovered
independently by \citet{Udry03} and \citet{Butler03}. At that time, its mass and orbital
parameters (e.g. period $P_c \sim 2000$~days) were highly uncertain. Finally,
\citet{Vogt05} announced discovery of the third planet. Its orbit was most likely located
between the orbits of planets `b' and `c'.\footnote{There is no clear consensus between
researchers about notation of planets in the system. The innermost planet is always
denoted as `b', but the notation for the outer pair of planets may vary. We use the same
notation as used in \emph{The Extrasolar Planets Encyclopaedia} by J.~Schneider, {\tt
www.exoplanet.eu}. Namely, we denote the innermost, the outermost and the intermediate
planet by the letters `b',`c',`d', respectively.}

Still, the radial velocity (RV) data are insufficient to obtain reliable estimations of
parameters of this system directly. Any attempt to obtain a best-fitting RV solution
inevitably leads to dynamically unstable orbital configuration disintegrating after a very
short time due to high eccentricities of two outer planets. To force the fitting algorithm
to find a stable configuration, \citet{Vogt05} fixed the value of $e_c$ at $0.2$.
\citet{Gozd06,Gozd08} presented a detailed analysis of the Keck RV data involving
constraints of dynamical stability. The stable orbital configurations from these works
span a very wide region, with ratio of orbital periods of the outer planets, $P_c/P_d$,
ranging from $\sim 2.1$ to $\sim 2.9$. We have to ascertain that, in fact, the Keck RV
data only outline a wide region of acceptable orbital configurations, whereas significant
constraints on the orbits of these planets are set mainly by the stability requirement.

The aim of the present paper is to describe the set of most likely orbital solutions for
the system of HD37124, based on the analysis of the complete set of RV data published
(incorporating Keck, ELODIE, and CORALIE measurements). The structure of the paper is as
follows. In Section~\ref{sec_data}, the RV datasets used in the paper are described. In
Section~\ref{sec_analysis}, the statistical methods used in the paper are discussed. In
Sections~\ref{sec_bestfit}, \ref{sec_dynam}, and~\ref{sec_ACR}, the results of the
analysis are presented. In Section~\ref{sec_evol}, the dynamical behaviour of the
resulting orbital configurations is considered. In Section~\ref{sec_extra}, the hypothesis
of existence of an extra planet in the system is tested.

\section{Radial velocity data}
\label{sec_data}
\begin{figure}\sidecaption
\begin{tabular}{@{}c@{}}
 \includegraphics[width=0.70\linewidth]{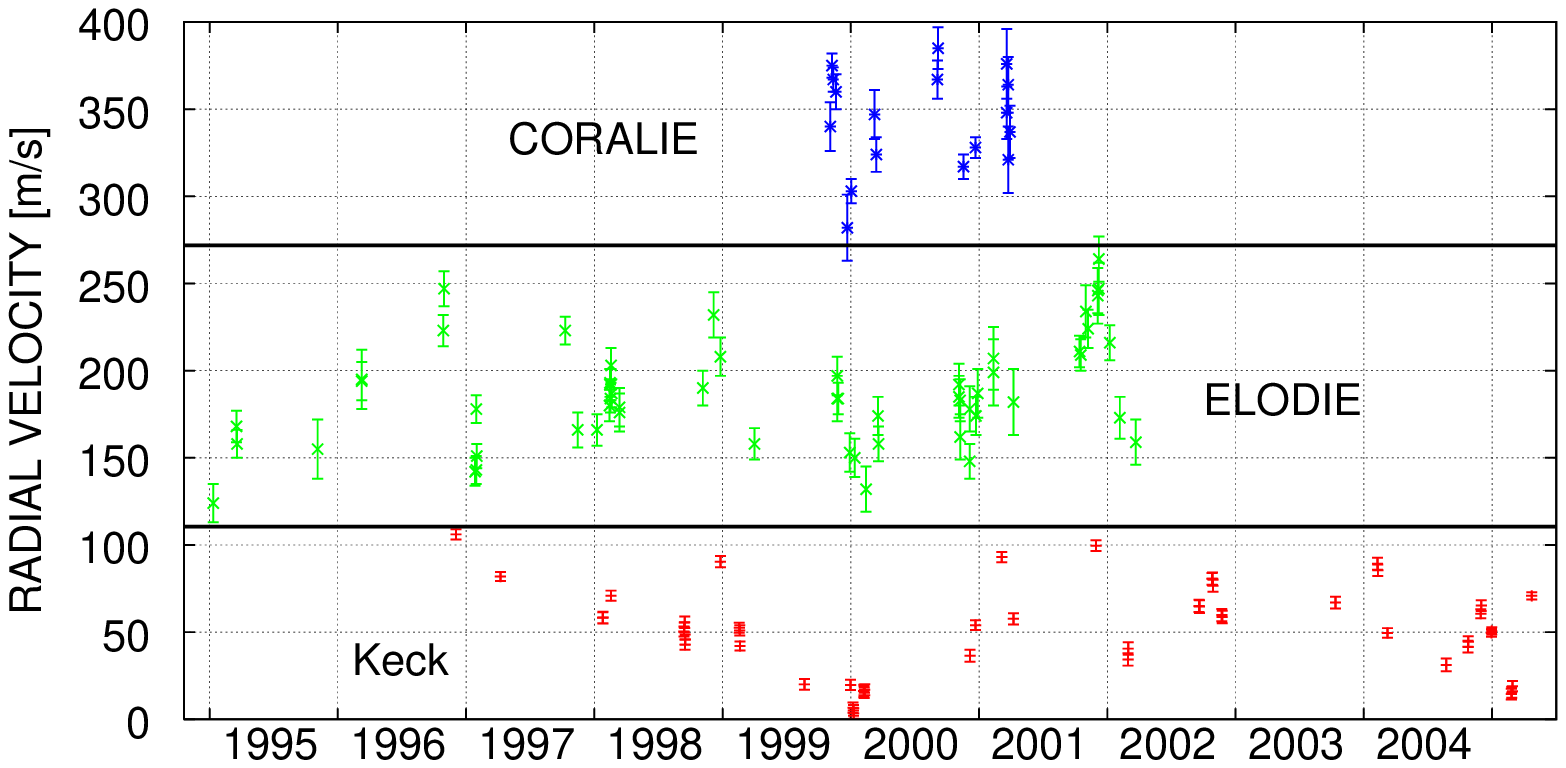} \\
 \includegraphics[width=0.70\linewidth]{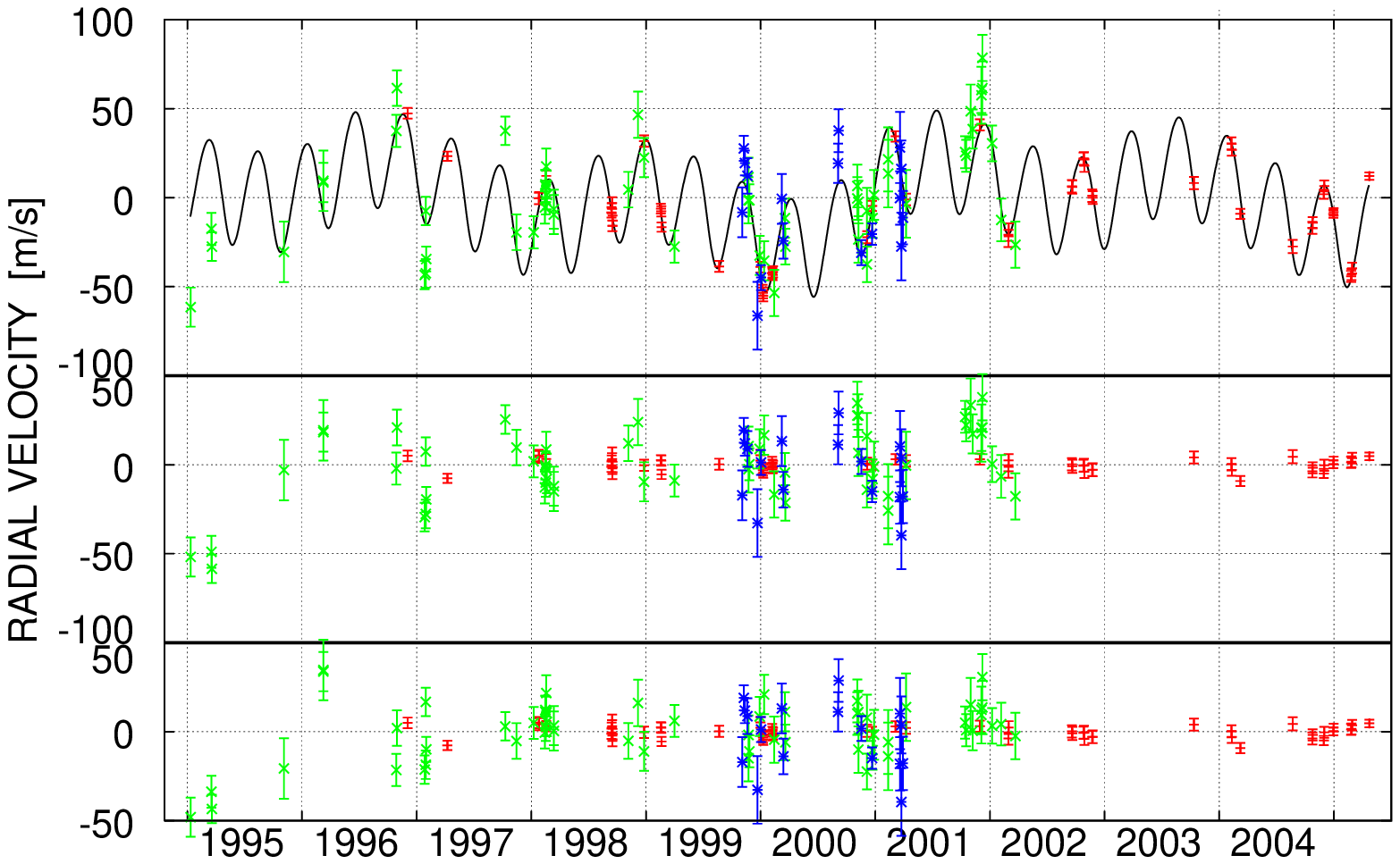}
\end{tabular}
\caption{Top panel: RV measurements for the star HD37124. Different RV offsets were
assigned to the ELODIE, CORALIE, and Keck datasets, in order to separate them from each
other. Bottom panel: RV curve for the best stable fit from \citep{Gozd08}, its RV
residuals, and its RV residuals after correction of the best-fitting sinusoidal annual
drift in the ELODIE data.}
\label{fig_RV}
\end{figure}

The most precise publicly available radial velocity data for HD37124 were published in the
paper by \citet{Vogt05}. These $52$ measurements were obtained at the Keck telescope and
span about $8.4$~yr (between $1996.9$ and $2005.3$). They show RV uncertainties from
$2.1$~m/s to $3.7$~m/s. Also, the observations at ELODIE and CORALIE instruments were
made. Although these are not so precise, they could significantly increase the temporal
coverage of the full RV time series. Unfortunately, these data were not published in a
table form and only a graph of these measurements is available in the paper
by~\citet{Udry03}. We apply a similar approach as \citet{FerrazMello05} and
\citet{Beauge08} used for HD82943. We reconstruct the measurements, their uncertainties,
and their dates from the graph published in~\citep{Udry03}. The radial velocities
themselves and their error bars can be reconstructed quite accurately (better than $1$~m/s
accuracy). The reconstructed Julian dates of observations have typical errors of $\sim
1$~day. This is admissible also, because the shortest orbital period $P_b \approx
150$~days in HD37124 is much longer. The $58$ reconstructed ELODIE data points cover about
$7.2$~yr between $1995.0$ and $2002.2$ and possess RV uncertainties from $7$~m/s to
$19$~m/s. The $17$ reconstructed CORALIE data points cover about $1.4$~yr between $1999.8$
and $2001.2$ and possess RV uncertainties from $6$~m/s to $20$~m/s. Thus the span of the
combined time series is about $10.3$~yr. This combined time series incorporate $J=3$
independent time series of $N_j (j=1,2,3)$ RV measurements $v_{ji} (i=1,2,\ldots,N_j)$
having the `stated' RV uncertainties $\sigma_{\mathrm{meas},ji}$ and made at the epochs
$t_{ji}$. These datasets are plotted in Fig.~\ref{fig_RV}, top panel.

It was demonstrated in \citep{Baluev08-IAUS249,Baluev08b} that high-precision RV
measurements in planet search surveys often suffer from periodic (annual) systematic
errors which may originate from various sources. Sometimes, these systematic errors may
reach the magnitude $\sim 10$~m/s, especially for data published several years ago, when
data reduction algorithms have not been debugged to a perfect state yet. It is necessary
to account for these systematic errors in our analysis. For this purpose, a simple
harmonic model of these errors $A \cos(2\pi(t-\tau)/1\mathrm{yr})$ is adopted below. Here,
the semi-amplitude $A$ and the time shift $\tau$ are the extra free parameters to be
determined from the time series. In the next section we will see that ELODIE data always
show a significant annual drift of radial velocity with large semi-amplitude $A \sim
20$~m/s. Unfortunately, the number of CORALIE measurements is too small for reliable
modelling of their possible annual errors. Seemingly, it is better to avoid this modelling
for the CORALIE data. The existence of significant annual errors in Keck measurements is
uncertain. We will consider different models of the RV curve below, with and without the
annual term in the Keck data.

One might ask that since the data from ELODIE and CORALIE are less accurate, suffer from
significant systematic errors, and increase the total time coverage by only $20-25\%$, why
to include them into analysis at all? If the orbits of planets in the system were
determined well by the Keck data alone then indeed addition of such RV data would not
significantly improve the precision of the estimations. However, the Keck data alone
provide only quite mild constraints on the orbits of the two outer planet. The major
problems are due to ill-determinacy of the orbital period and eccentricity of the
outermost planet. In this case, we should compare RV datasets in the sense of absolute
rather than relative increase of the total time base. About ten ELODIE data points, which
span two years before the regular observations of HD37124 started at the Keck observatory,
cover almost full orbital period of the intermediate planet and about one third or almost
half of the orbital period of the outermost planet. During this time segment, the
corresponding RV variations are about $\sim 30$~m/s for the planet d and $\sim 10-20$~m/s
for the planet c. Such arcs of the RV curves are very important for constraining the
corresponding orbital periods, which result in more strict constraining of the whole set
of parameters. Of course, the Keck data remain the main source of information and drive
the fit, but the data from ELODIE are also important, because they can help to rule out a
large fraction of inacceptable fits. This is illustrated by the bottom panel in
Fig.~\ref{fig_RV}. We can see that the orbital solution from the work \citep{Gozd08} fits
satisfactorily all available RV data in the range 1997--2005, except for the ELODIE
measurements made in 1995--1996, which show systematic deviation reaching $50$~m/s. This
makes the mentioned orbital solution significantly less credible. Other solutions, not
being ruled out by the Keck data, may produce even larger deviations and thus could be
easily ruled out even by very inaccurate measurements. However, data from different
observatories have different statistical properties and it is necessary to merge them
extremely carefully, in order to set correct statistical weights to different datasets.
This problem will be considered in the next section.

\section{Statistical analysis: principles and definitions}
\label{sec_analysis}
\subsection{The general RV model and the system of parameters}
\label{sec_model}
Let us write down the model of the radial velocity measurements obtained at $j^{\rm th}$
observatory at time $t$ as
\begin{equation}
 \mu_j(t,\vec p) = \mu_{\mathrm{obs},j}(t,\vec p_{\mathrm{obs},j}) +
                      \mu_\star(t,\vec p_\star), \qquad j = 1,2,\ldots J.
\label{RVfull}
\end{equation}
Here, the full vector $\vec p$ of free parameters to be estimated consists of elements
of vectors $\vec p_{\mathrm{obs},j}\, (j=1,2,\ldots, J)$ and $\vec p_\star$. The
function $\mu_{\mathrm{obs},j}$ in~(\ref{RVfull}) represent an observatory-specific part
of the measured radial velocity:
\begin{equation}
 \mu_{\mathrm{obs},j}(t,\vec p_{\mathrm{obs},j}) = c_{0,j} +
            \sum_{n=1}^{s_j} A_{jn} \cos(2\pi(t-\tau_{jn})/P_{jn})
\label{RVobs}
\end{equation}
The constant velocity term $c_{0,j}$ and parameters $A_{jn},\tau_{jn}$ of possible
systematic errors form the vectors $\vec p_{\mathrm{obs},j}$ of unknowns. The
quantities $P_{jn}$~--- the periods of the systematic errors~--- are assumed \emph{a
priori} known. In this paper, we will consider only the cases $s_j=0$ (no systematic
errors) and $s_j=1$ with $P_{j1}$ being the annual period. The function $\mu_\star$
in~(\ref{RVfull}) is the common radial velocity term incorporating RV signals due to
unseen companions orbiting the star:
\begin{equation}
 \mu_\star(t,\vec p_\star) = \sum_{n=1}^r c_n t^n +
             \sum_{n=1}^{\mathcal N} K_n (\cos(\omega_n+\upsilon_n) + e_n\cos\omega_n).
\label{RVplanets}
\end{equation}
The coefficients $c_n$ describe possible long-term polynomial (of degree $r$ in general,
we will consider only the cases $r=0$, no trend, and $r=1$, linear trend, below) trend in
the RV data. This trend may be induced by possible distant unseen companions in the system
with periods longer than the time span of the observations. Other terms
in~(\ref{RVplanets}) represent Keplerian velocities induced by $\mathcal N$ planets
($\mathcal N=3$ in our case). The coefficients $c_n$, RV semi-amplitudes $K_n$ and the
orbital elements $\lambda_n$ (the mean longitude at certain fixed epoch), $e_n$ (the
eccentricity), $\omega_n$ (the argument of the periastron), $P_n$ (the orbital period)
form the vector $\vec p_\star$ of planetary parameters to be estimated. The quantity
$\upsilon_n$ in~(\ref{RVplanets}) is the true anomaly of the $n^{\rm th}$ planet
(evidently, it depends on the time and on the parameters $\lambda_n,P_n,e_n$).

\begin{figure}\sidecaption
\includegraphics[width=0.55\textwidth]{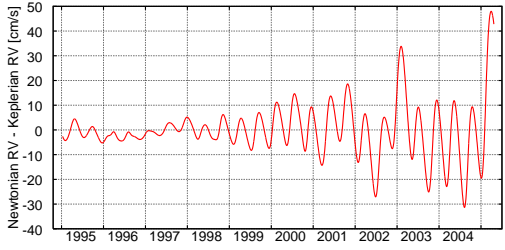}
\caption{The difference between the Newtonian N-body and multi-Keplerian RV models for the
fit from \citep{Gozd08}. The clear growth of this difference up to $\sim 50$~cm/s reflects
the fact that the orbital elements were given for the epoch of the first Keck observation.
If the reference epoch was closer to the middle of the observation window, the RV
difference would be bounded by $20-30$~cm/s over the whole time segment.}
\label{fig_RVdiff}
\end{figure}

The model~(\ref{RVplanets}) does not take into account interactions between the planets.
This is admissible for the case of HD37124, because after $\sim 10$~yr of RV observations,
the outermost planets `c' and `d' (which, as we will see below, represent the main source
of dynamical activity in the system) have completed only two and four revolutions around
the star. On such a time scale, the planetary perturbations could be only significant in
the case of sufficiently close approaches between planets on high-eccentricity orbits.
However, such configurations are unrealistic, because close approaches usually represent
the source of instability and lead the system to disintegration. For example, the
difference between the best-fitting Keplerian and N-body RV models (see
Fig.~\ref{fig_RVdiff}) do not exceed $50$~cm/s for the best \emph{stable} orbital
configuration from \citep{Gozd08}. For \emph{stable} orbital solutions that are considered
below, these deviations do not exceed $\sim 30$~cm/s and produce only $\sim 10^{-3}$
relative error in the RV residuals r.m.s. Bearing this in mind, only the Keplerian RV
models are used in the paper. They allow much faster computations than the N-body ones.

The minimum mass of the planet $m\sin i$ (here $i$ is the orbital inclination to the sky
plane) and its semi-major axis $a$ can be derived as
\begin{eqnarray}
 m \sin i &\simeq& \tilde K \left( \frac{M_\star^2 P}{2\pi G}\right)^{1/3} =
            \mathcal M \tilde K P^{1/3} M_\star^{2/3}, \nonumber\\
 a &\simeq& \left( \frac{G M_\star P^2}{4\pi^2} \right)^{1/3} = \mathcal A P^{2/3} M_\star^{1/3},
\label{mass-SMA}
\end{eqnarray}
where $\mathcal M\approx 4.919\cdot 10^{-3}$~[$M_{Jup}\cdot M_\odot^{-2/3}\cdot {\rm
m}^{-1}\cdot {\rm day}^{-1/3}\cdot {\rm s}$] and $\mathcal A \approx 1.957\cdot
10^{-2}$~[${\rm AU} \cdot M_\odot^{-1/3}\cdot {\rm day}^{-2/3}$] are constant factors, $G$
is the gravitational constant, $M_\star$ is the mass of the star, and $\tilde K =
K\sqrt{1-e^2}$ is the modified semi-amplitude. As it is well known, the inclination $i$
cannot be constrained using the Keplerian RV model.

The errors of the approximate equalities in~(\ref{mass-SMA}) are about $m/M_\star
\sim 10^{-3}$ and are much less than the statistical uncertainties of estimations given
below. For example, the shift in the RV semi-amplitude of about $m/M_\star \sim 10^{-3}$
would lead to an error in the RV of only $1-3$~cm/s and would be completely invisible in
the resulting fit. Regular motions are not sensitive to such small shifts in the planetary
masses. If the planets are not trapped in a mean-motion resonance, the same proposition
holds true for the semi-major axes as well. However, we deal below with resonant motions,
which are sensitive to the shifts of the semi-major axes of $\mathcal O(m/M_\star)$. For
these cases, we need to provide more decimal digits (excessive with respect to the
statistical uncertainties) for the values of the semi-major axes, in order to allow the
reader to reproduce the results of long-term numerical integrations discussed below. For
this purpose we should state the coordinate system in which we refer our estimations. It
was noted by \citet{LissRiv01} and \citet{LeePeale03} that it is better to interpret the
orbital parameters of the Keplerian model as osculating ones referenced in the Jacobi
coordinate system. In the Jacobi coordinates, the osculating orbital period and semi-major
axis of an $k^{\rm th}$ planet are connected by the relation
\begin{equation}
a_k = \mathcal A P_k^{2/3} \left(M_\star+\sum_{j=1}^{k} m_j\right)^{1/3}.
\end{equation}
Here we need to substitute the values of the planetary masses themselves, and thus to
assume some values for orbital inclinations. As it will be discussed in
Section~\ref{sec_dynam}, we assume that the planetary system is seen edge-on, that is
$i_k=90^\circ$.

\subsection{Estimations of the parameters}
\label{sec_estim}
To analyse the RV data described in Section~\ref{sec_data}, we need to estimate the
so-called RV jitter $\sigma_\star^2$ which increases the full RV uncertainties as
$\sigma_{\mathrm{full}}^2 = \sigma_\star^2 + \sigma_{\mathrm{meas}}^2$ and softens the
differences between the weights of observations $\propto 1/\sigma_{\mathrm{full}}^2$. In
the astrophysical part, this RV jitter is inspired by various activity in the star
\citep[e.g.][]{Wright05}, but often incorporates instrumental effects as well. As is shown
by \citet{Baluev08b}, the effective RV jitter may be quite different for different
instruments, even for one and the same star. Therefore, we should perform the merging of
RV datasets from different observatories very carefully, with correct assignment of
statistical weights to these data. To do it, we use here the maximum-likelihood approach
described in the paper \citep{Baluev08b}. This algorithm includes a built-in estimation of
the effective RV jitter (simultaneous with the estimation of usual parameters), which
allows us not to rely on a low-precision astrophysical estimations of $\sigma_\star$.
Moreover, this algorithm allows to perform a separate estimation of the effective RV
jitters for the datasets from different observatories. This algorithm uses the
maximization of the modified log-likelihood function of the $N$ RV observations $v_{ji}$
(their errors are assumed to be uncorrelated and Gaussian-distributed) which is defined as
\begin{equation}
 \ln \tilde{\mathcal L} = - \sum_{j=1}^J \sum_{i=1}^{N_j} \left[ \frac{\left( v_{ji} -
      \mu_j(t_{ji},\vec p) \right)^2}{2\gamma \sigma_{\mathrm{full},ji}^2} + \ln
      \sigma_{\mathrm{full},ji} \right] - N \ln\sqrt{2\pi}
      \,\stackrel{\sigma_\star^2,\vec p}{\longrightarrow}\, \max.
\label{likmod}
\end{equation}
Here, $\sigma_{\mathrm{full},ji}^2 = \sigma_{\star,j}^2 + \sigma_{\mathrm{meas},ji}^2$
($j=1,2,\ldots,J; i=1,2,\ldots N_j$) and the correction divisor $\gamma = 1 - d/N$ with
$d$ being the number of degrees of freedom in our RV model (i.e., the number of free
parameters, $d = \dim \vec p$). The divisor $\gamma$ allows to perform a `preventive'
reduction of the statistical bias in the estimations of the RV jitter. Since we aim to use
the new objective function $\tilde{\mathcal L}$, instead of the usually used $\chi^2$ one,
we need to introduce a new measure of the goodness-of-fit, which would be based on
$\tilde{\mathcal L}$. In accordance with \citet{Baluev08b}, to assess the quality of a
given orbital fit, the following goodness-of-fit measure is used below:
\begin{equation}
 \tilde l = \tilde{\mathcal L}^{-1/N} e^{-0.5}/\sqrt{2\pi} \approx 0.2420 \tilde{\mathcal
 L}^{-1/N}.
\label{gofmod}
\end{equation}
This function is measured in the same units as radial velocity (i.e., in m/s). It
characterises the overall scatter of RV measurements around the model. However, to allow a
comparison with previous works on HD37124, the traditional r.m.s. goodness-of-fit measure
is used below as well.

\subsection{Assessing the reliability of orbital fits}
\label{sec_rely}
It is not enough to find an RV fit with a small scatter of residuals. To interpret the
resulting estimations, we need to assess their reliability. It is shown by
\citet{Beauge08} that orbital fits of multi-planetary systems in a mean-motion resonance
(hereafter MMR) may be highly unreliable, though formal uncertainties of estimations may
be apparently small. In these cases, the shape of the likelihood function may be
complicated and may possess multiple comparable local maxima. Often this shape is
model-dependent: addition of extra model components leads to qualitative changes of the
set of likelihood maxima. Often, every such local maximum provides a good fit of the RV
curve but nevertheless is far from the actual orbital configuration. Such situation
indicates one of the following items:
\begin{enumerate}
\item The adopted model is imperfect. Some extra terms were not included or the terms
included are wrong.
\item The data are imperfect. The errors may have a non-Gaussian distribution, they may be
correlated or they may incorporate some extra time-variable systematic part. Also, the
data may cover too small time base or simply the number of observations is too small.
\end{enumerate}
Formally, we could use one of the local maxima (e.g. the global maximum) to construct an
estimation of the parameters of the system. However, such estimation would appear strongly
biased (usually to higher eccentricities) and its formal uncertainties would strongly
underestimate real errors of parameters. Therefore, in addition to the formal
goodness-of-fit measure, we need some indicator of statistical reliability of our fits.

\citet{Beauge08} performed (for the system of HD82943) several fits with truncated
RV datasets to explore the sensitivity of current orbital fits to future RV measurements.
Unfortunately, this approach require too time-consuming computations. Here we need some
simple and rapid (though perhaps quite rough) test of the `statistical health' of our
orbital fits. For this goal, we use below the following approach. Recall
\citep[\S~6.4]{Lehman-est} that the asymptotic ($N\to\infty$) approximation to the
variance-covariance matrix of the estimations of $\vec p$ is calculated as the inverse
of the Fisher information matrix $\tens Q$ having elements
\begin{equation}
Q_{\alpha\beta} = \sum_{j=1}^J \sum_{i=1}^{N_j} \frac{1}{\sigma_{\mathrm{full},ji}^2}
            \left.\frac{\partial \mu_j}{\partial p_\alpha}\right|_{t=t_{ji}}
            \left.\frac{\partial \mu_j}{\partial p_\beta}\right|_{t=t_{ji}}.
\label{norm_matr}
\end{equation}
When we deal with a well-conditioned situation, the likelihood function can be
quadratically approximated in the vicinity of its maximum using the quadratic term
$\propto \delta \vec p^{\rm T} \tens Q \vec \delta \vec p$ only. This is the
asymptotic behaviour of the likelihood function for large time series (i.e.,
$N\to\infty$). The mentioned quadratic term approximates the multidimensional graph of the
likelihood function by a paraboloidal hypersurface. Other terms in the expansion of the
likelihood function are insignificant and do not distort this shape essentially. However,
when the number of observations is not sufficient for reliable determination of the
parameters of the model, the matrix $\tens Q$ become ill-conditioned, and the extra
terms easily produce distortions leading to multiple local maxima of the likelihood
function.

To assess the reliability of a particular fit we could calculate the condition number of
the information matrix $\tens Q$ (i.e. the ratio of the biggest eigenvalue and the
smallest one). The larger is this number, the lower is the reliability of the
corresponding orbital fit. This condition number should be compared with the number of
observations, because high-order terms in the expansion of the likelihood function tend to
zero when $N$ grows. To demonstrate this, let us write down the full expression of the
Hessian matrix of the usual $\chi^2$ function (speaking more accurately, of the weighted
average of squared residuals):
\begin{equation}
H_{\alpha\beta} = - 2 Q_{\alpha\beta} + 2 \sum_{j=1}^J \sum_{i=1}^{N_j}
\frac{\mu_j-v_{ji}}{\sigma_{\mathrm{full},ji}^2} \left. \frac{\partial^2 \mu_j }{\partial
p_\alpha \partial p_\beta} \right|_{t=t_{ji}}.
\label{Hess}
\end{equation}
In this expression, the first term determines the asymptotic behaviour of the likelihood
function in the vicinity of its maximum. The second term (summation) reflects the
non-linearity of the RV model and contains the residuals $(v-\mu)$, which average
decreases as $\mathcal O(1/\sqrt N)$ when $N$ grows. This implies that the perturbing term
in~(\ref{Hess}) grows according to the law $\mathcal O(\sqrt N)$, whereas the Fisher
matrix grows according to $\mathcal O(N)$. Therefore, the relative magnitude of the second
term in~(\ref{Hess}) is about $1/\sqrt N$. Let us now transform $\tens Q$ to its diagonal
form with eigenvalues $E_1 \ldots E_d$ on the diagonal (we assume that $E_1$ is the
biggest eigenvalue and $E_d$ is the smallest one). We may expect that the typical
magnitude of the elements of the Fisher matrix is of the order of $\sqrt{E_1 E_d}$ and,
therefore, the typical magnitude of the second term in~(\ref{Hess}) is about $\sqrt{E_1
E_d}/\sqrt N$. The full matrix $\tens H$ is well approximated by its asymptotic
representation $-2\tens Q$, if relative deviations of the corresponding eigenvalues are
small. If this is so, the topology of the graph of the likelihood function should not be
very sensitive to the perturbing terms. We can expect that \emph{absolute} deviations of
the eigenvalues should have similar magnitude about $\sqrt{E_1 E_d / N}$. But the
\emph{relative} deviation of the smallest eigenvalue $E_d$ is as large as
$\sqrt{E_1/E_d}/\sqrt N$. Therefore, the condition number $E_1/E_d$ of the Fisher
information matrix should not exceed the number of observations $N$ for a statistically
robust orbital fit. This is simply a specification of the general fact, known from the
numerical analysis, that the condition number is a measure of sensitivity of a matrix to
small perturbations.

There is a small clause. The elements of $\tens Q$ are measured in different physical
units and it would be incorrect to use the condition number of $\tens Q$ itself (since it
would depend on the choice of the measurement units). Instead, we may use the condition
number $\mathcal C$ of the scaled information matrix $\tilde{\tens Q}$ having elements
$\tilde Q_{ij} = Q_{ij}/\sqrt{Q_{ii}Q_{jj}}$.

Unfortunately, the indicator $\mathcal C$ depends on the epoch for which we obtain the
fit, especially when the RV model incorporates long-term trends. To obtain an informative
estimation of $\mathcal C$, we should choose the reference epoch near the middle of the
time series span (or, when we merge datasets with different characteristics, near the
weighted average of the timings $t_{ij}$). In this position, the value of $\mathcal C$
will be close to its minimum and, simultaneously, the set of parameters will possess the
best statistical properties (e.g., the mutual correlations of different parameters will be
minimized). Even if absolute value of the quantity $\mathcal C$ represents a too rough
indicator of the fit robustness, this indicator seems to work quite well for the goal of
intercomparison of different orbital fits for the same planetary system. To illustrate the
use of the indicator $\mathcal C$, we give here its (minimum) values for several planetary
systems with well-determined orbits: $\mathcal C=10$ with $N=409$ (51Peg), $\mathcal
C=6.7$ with $N=109$ (70Vir), $\mathcal C=9.1$ with $N=203$ (14Her). Less perfect cases:
$\mathcal C=17$ with $N=75$ (HD69830), $\mathcal C=120$ with $N=487$ (55Cnc). For the
system of HD82943, we obtain $\mathcal C=200$ with $N=165$ (three-planet model with the
eccentricity of the outermost planet fixed at zero).

\section{Best-fitting orbital solutions for HD37124}
\label{sec_bestfit}
From now on, we consider three main models of the RV data for HD37124. All models
incorporate three common Keplerian terms as in the eq.~(\ref{RVplanets}), the constant
velocity terms (separate for the three datasets), and the annual term for the ELODIE
dataset. The first RV model (I) does not incorporate anything else and thus has $d=20$
degrees of freedom. The second one (II) incorporates also an annual term for the Keck
dataset and has $d=22$. The third one (III) incorporates the same terms as (I) plus a
linear trend (common for all the three datasets) and thus has $d=21$. For all these
models, the ratio $d/N \approx 0.16$ means that we are left with only $\approx 6$
observations per one parameter to be estimated. This indicates that the problem of
obtaining a suitable orbital configuration of the system cannot be solved easily.

Let us try to reconstruct the topological structure of the multidimensional graph of the
likelihood function~(\ref{likmod}). Since the dimension of the problem is large, we cannot
look on corresponding hypersurface directly. Let us pick two (of $d$) free parameters
$x,y$ (say the orbital periods $P_c$ and $P_d$ of the outer planets). Then we consider the
function $\tilde l'(x,y) = \min \tilde l$, where the minimization of the goodness-of-fit
function $\tilde l$ is performed over the rest of free parameters. This means that for any
manually assigned values of $x,y$ we obtain corresponding best-fitting values of other
parameters and find corresponding goodness-of-fit measure $\tilde l$. Further, the
resulting function of two variables can be visualised on a two-dimensional grid.

\begin{figure}
\begin{tabular}{@{}c@{}c@{}c@{}}
 \includegraphics[width=0.34\linewidth]{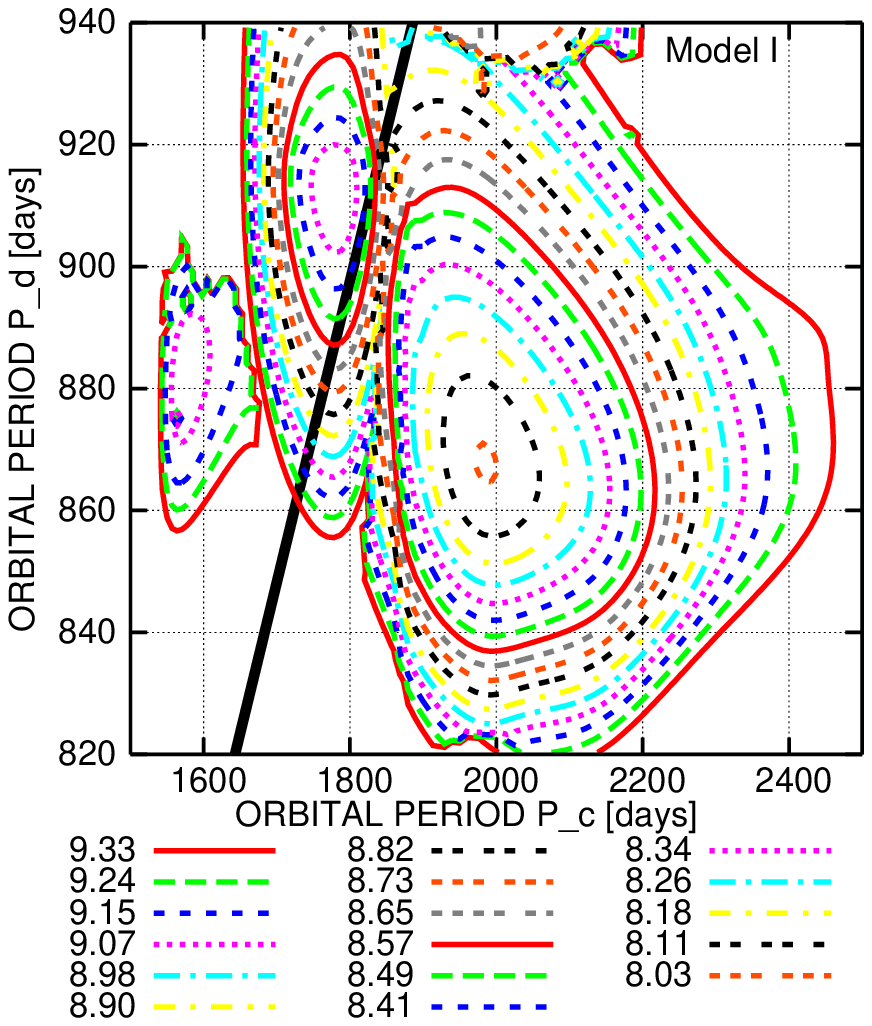} &
 \includegraphics[width=0.32\linewidth]{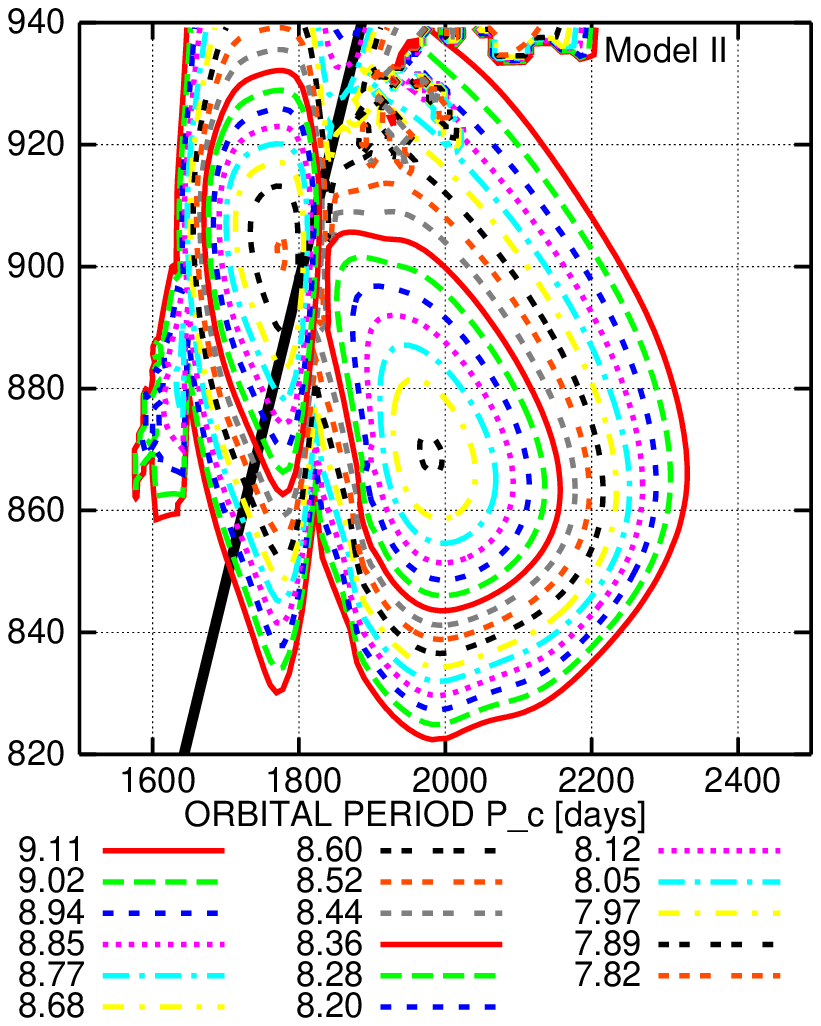} &
 \includegraphics[width=0.32\linewidth]{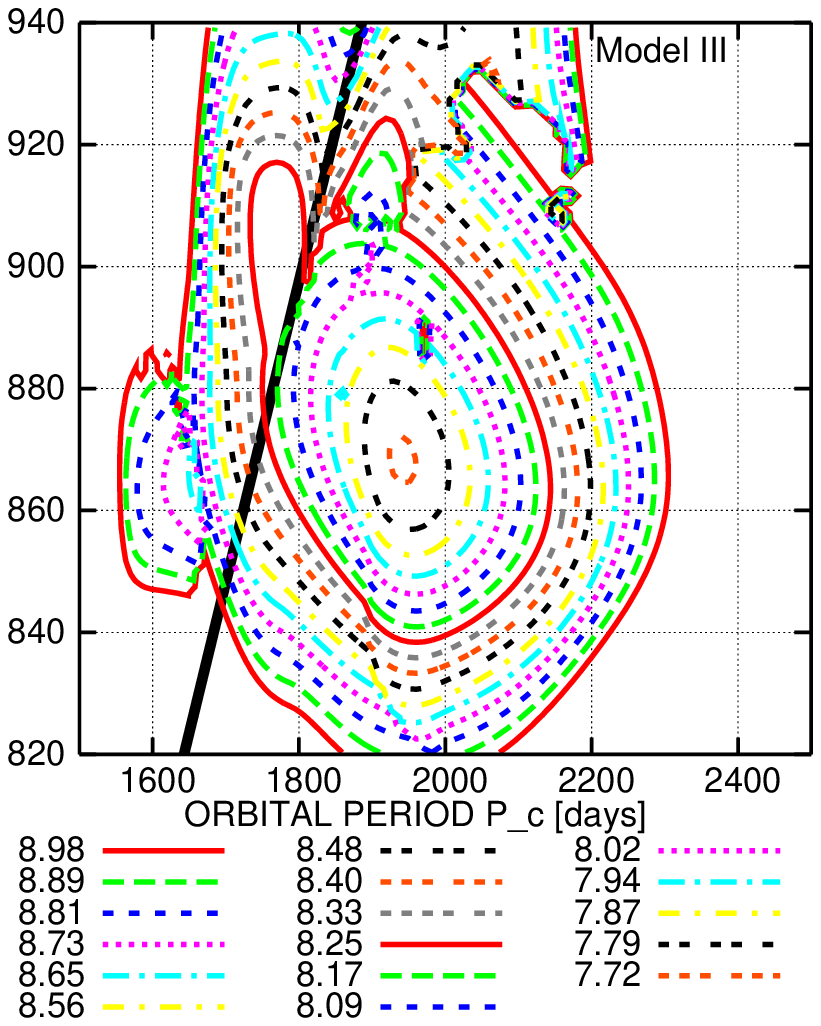} \\
 \includegraphics[width=0.34\linewidth]{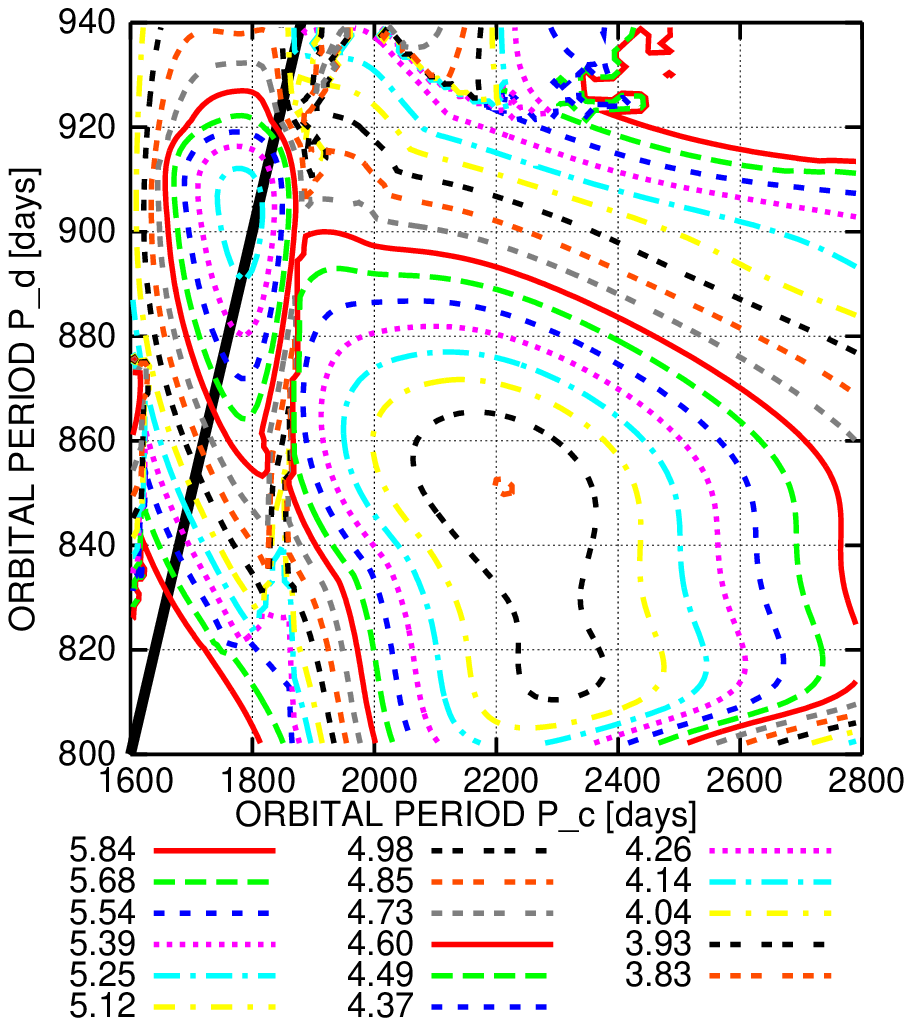} &
 \includegraphics[width=0.32\linewidth]{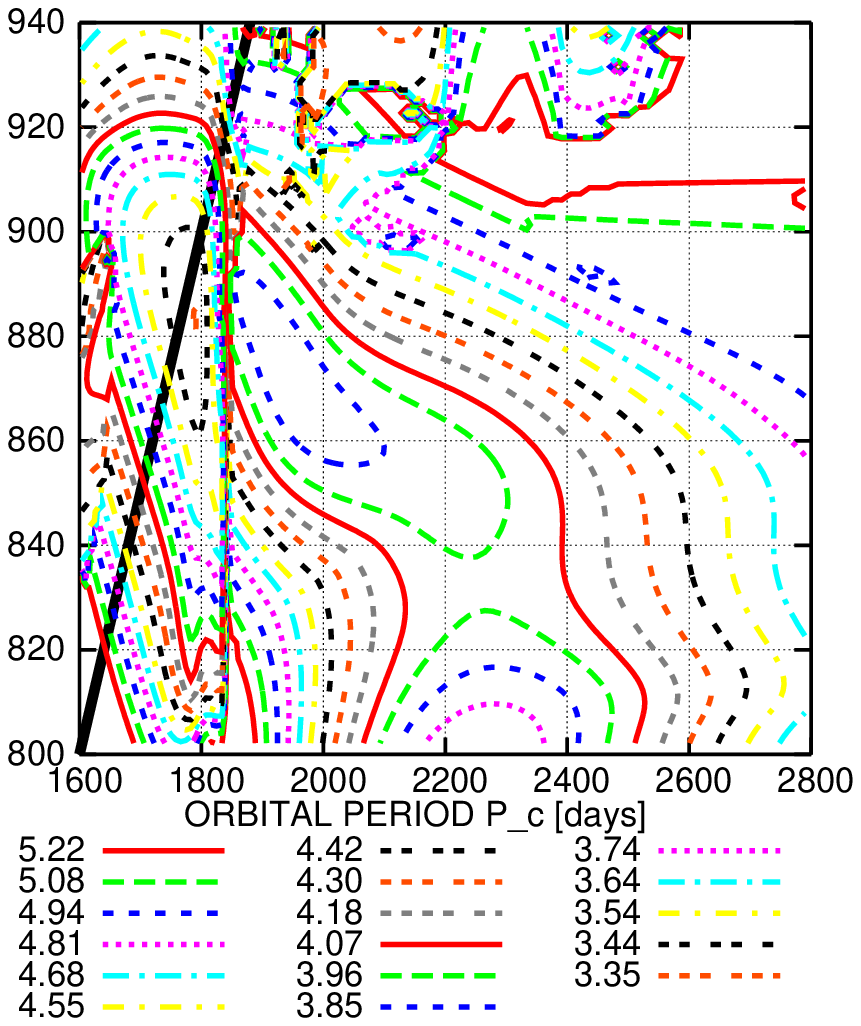} &
 \includegraphics[width=0.32\linewidth]{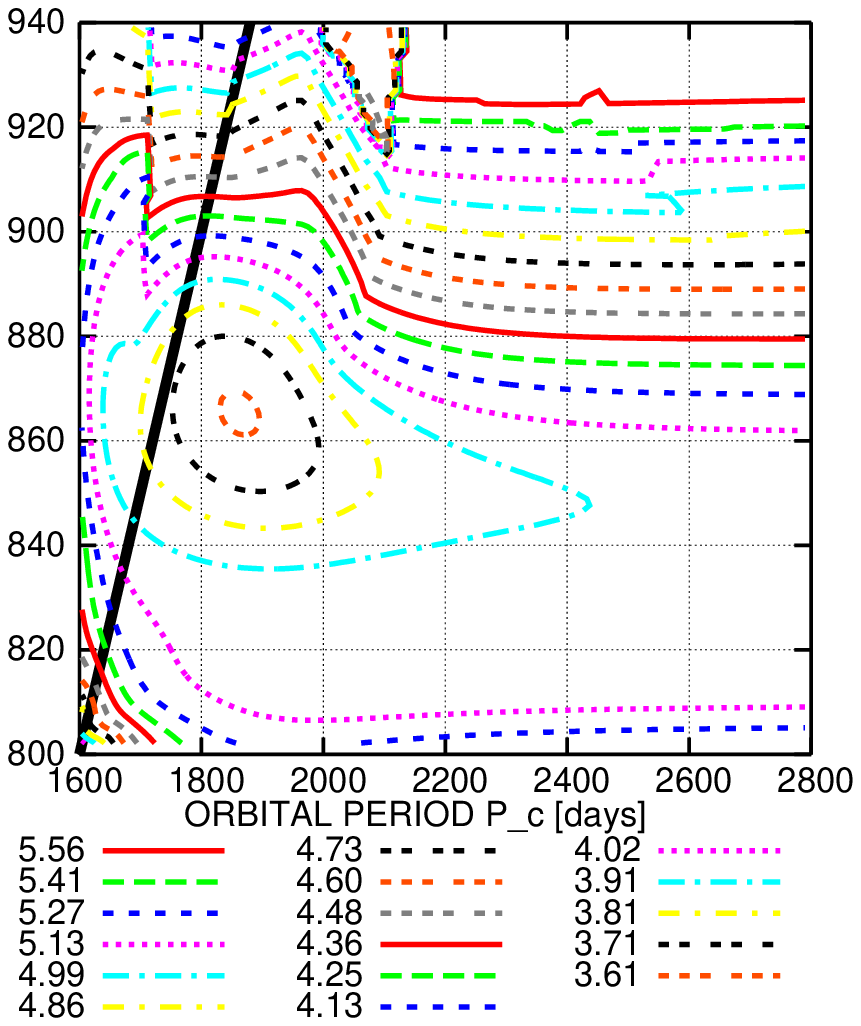} \\
\end{tabular}
\caption{Contour maps of the likelihood goodness-of-fit statistic $\min \tilde l$ for
RV fits of HD37124. The maps are plotted in the plane of orbital periods $P_c$ and $P_d$
(measured in days). For each point in these panels, the likelihood goodness-of-fit
function $\tilde l$ was minimized over the rest of free parameters. Thick straight line
mark the position of the 2/1 MMR. The level values of the plotted function are shown below
the respective panels. The panels in different columns correspond to the different RV
models (model~I, model~II, model~III, from left to right). The top raw of panels
corresponds to the fits based on the whole accessible data array, whereas the bottom raw
corresponds to the fits based on the Keck data only. Note the differences in period ranges
shown in top and bottom graphs.}
\label{fig_Pc-Pd}
\end{figure}

\begin{table}
\caption{Best-fitting orbital solutions for the planetary system around HD~37124.}
\begin{tabular}{@{}lllllll@{}}
\hline\noalign{\smallskip}
parameter            & IA, 2/1         & IB              & IIA, 2/1        & IIB             & IIIA, 2/1       & IIIB            \\
\noalign{\smallskip}\hline\noalign{\smallskip}
\multicolumn{7}{c}{planet b}\\
$P$~[days]           & $154.37(13)$    & $154.48(12)$    & $154.53(12)$    & $154.58(12)$    & $154.37(13)$    & $154.49(12)$    \\
$\tilde K$~[m/s]     & $28.39(82)$     & $28.49(85)$     & $27.42(73)$     & $28.39(85)$     & $28.06(83)$     & $27.98(81)$     \\
$\lambda$~[$^o$]     & $117.7(1.9)$    & $119.1(1.8)$    & $118.2(1.8)$    & $118.8(1.7)$    & $118.0(1.8)$    & $120.2(1.7)$    \\
$e$                  & $0.116(33)$     & $0.116(33)$     & $0.108(29)$     & $0.081(27)$     & $0.093(33)$     & $0.064(28)$     \\
$\omega$~[$^o$]      & $131(15)$       & $132(23)$       & $134(14)$       & $132(23)$       & $135(18)$       & $140(25)$       \\
$m\sin i$~[$M_{Jup}$]& $0.635(46)$     & $0.637(47)$     & $0.613(44)$     & $0.635(46)$     & $0.628(46)$     & $0.626(46)$     \\
$a$~[AU]             & $0.518(17)$     & $0.519(17)$     & $0.519(17)$     & $0.518(17)$     & $0.518(17)$     & $0.518(17)$     \\
\multicolumn{7}{c}{planet d}\\
$P$~[days]           & $912.5(7.7)$    & $867.7(9.1)$    & $902.3(8.7)$    & $869.3(9.5)$    & $904(14)$       & $868.3(8.5)$    \\
$\tilde K$~[m/s]     & $13.8(1.2)$     & $13.3(1.0)$     & $15.1(1.0)$     & $13.70(97)$     & $13.1(1.1)$     & $14.1(1.0)$     \\
$\lambda$~[$^o$]     & $290.3(7.4)$    & $345.9(4.3)$    & $289.5(5.4)$    & $344.7(4.2)$    & $297.8(8.3)$    & $346.0(4.0)$    \\
$e$                  & $0.505(88)$     & $0.056(72)$     & $0.441(65)$     & $0.207(87)$     & $0.39(10)$      & $0.025(68)$     \\
$\omega$~[$^o$]      & $283(12)$       & $75(77)$        & $276(13)$       & $49(19)$        & $288(17)$       & $390(150)$      \\
$m\sin i$~[$M_{Jup}$]& $0.559(61)$     & $0.528(54)$     & $0.609(60)$     & $0.545(53)$     & $0.527(58)$     & $0.560(55)$     \\
$a$~[AU]             & $1.695(57)$     & $1.639(56)$     & $1.682(57)$     & $1.641(56)$     & $1.684(59)$     & $1.639(56)$     \\
\multicolumn{7}{c}{planet c}\\
$P$~[days]           & $1782(41)$      & $1985(50)$      & $1777(24)$      & $1980(55)$      & $1776(41)$      & $1940(56)$      \\
$\tilde K$~[m/s]     & $14.9(1.1)$     & $17.1(2.5)$     & $15.2(1.0)$     & $17.5(2.4)$     & $14.2(1.2)$     & $16.7(2.4)$     \\
$\lambda$~[$^o$]     & $319.8(6.1)$    & $309.3(5.2)$    & $322.5(4.5)$    & $312.7(4.6)$    & $311.9(7.1)$    & $306.2(5.7)$    \\
$e$                  & $0.527(70)$     & $0.593(85)$     & $0.563(56)$     & $0.555(72)$     & $0.467(83)$     & $0.573(84)$     \\
$\omega$~[$^o$]      & $256.2(9.6)$    & $325.8(7.6)$    & $251.9(6.6)$    & $335.0(8.2)$    & $247(13)$       & $320.9(8.7)$    \\
$m\sin i$~[$M_{Jup}$]& $0.754(77)$     & $0.895(61)$     & $0.766(72)$     & $0.915(62)$     & $0.720(76)$     & $0.867(59)$     \\
$a$~[AU]             & $2.648(97)$     & $2.85(11)$      & $2.643(91)$     & $2.94(11)$      & $2.642(97)$     & $2.80(11)$      \\
\noalign{\smallskip}\hline\noalign{\smallskip}
$c_1$~[$\frac{\rm m}{\rm s \cdot \rm yr}$]%
                     &     -           &    -            &     -           &     -           & $0.58(33)$      & $0.78(27)$      \\
\noalign{\smallskip}\hline\noalign{\smallskip}
\multicolumn{7}{c}{ELODIE dataset}\\
$c_0$~[m/s]          & $65.5(4.3)$     & $62.7(4.2)$     & $66.5(4.5)$     & $61.2(4.5)$     & $67.4(4.4)$     & $65.6(4.3)$     \\
$A$~[m/s]            & $17.6(3.2)$     & $17.8(2.6)$     & $17.5(3.5)$     & $16.9(2.6)$     & $18.2(3.5)$     & $17.2(2.9)$     \\
$\tau$~[days]        & $167(20)$       & $178(20)$       & $164(20)$       & $-181(22)$      & $162(18)$       & $171(20)$       \\
$\sigma_\star$~[m/s] & $6.6(2.3)$      & $5.9(2.3)$      & $7.4(2.3)$      & $6.5(2.3)$      & $6.3(2.3)$      & $5.6(2.4)$      \\
r.m.s.~[m/s]         & $11.85$         & $10.96$         & $12.23$         & $11.21$         & $11.79$         & $10.45$         \\
\multicolumn{7}{c}{CORALIE dataset}\\
$c_0$~[m/s]          & $4.3(4.2)$      & $7.9(4.2)$      & $4.8(4.3)$      & $7.4(4.2)$      & $5.3(4.4)$      & $8.6(4.4)$      \\
$\sigma_\star$~[m/s] & $11.3(4.0)$     & $10.9(4.0)$     & $12.3(4.2)$     & $11.2(4.1)$     & $12.1(4.1)$     & $12.0(4.1)$     \\
r.m.s.~[m/s]         & $17.45$         & $16.93$         & $17.62$         & $16.73$         & $17.68$         & $17.38$         \\
\multicolumn{7}{c}{Keck dataset}\\
$c_0$~[m/s]          & $7.56(90)$      & $9.11(84)$      & $6.8(1.0)$      & $10.8(1.1)$     & $7.47(80)$      & $9.00(79)$      \\
$A$~[m/s]            &     -           &    -            & $4.5(1.4)$      & $3.8(1.6)$      &     -           &    -            \\
$\tau$~[days]        &     -           &    -            & $-148(19)$      & $102(14)$       &     -           &    -            \\
$\sigma_\star$~[m/s] & $2.58(54)$      & $2.48(55)$      & $1.24(71)$      & $2.02(56)$      & $2.44(54)$      & $2.01(57)$      \\
r.m.s.~[m/s]         & $3.704$         & $3.713$         & $3.044$         & $3.381$         & $3.566$         & $3.359$         \\
\noalign{\smallskip}\hline\noalign{\smallskip}
$d$                  & \multicolumn{2}{c}{$20$}          & \multicolumn{2}{c}{$22$}          & \multicolumn{2}{c}{$21$}          \\
$\tilde l$~[m/s]     & $8.269$         & $8.025$         & $7.816$         & $7.888$         & $8.180$         & $7.711$         \\
$\mathcal C$         & $213$           & $54$            & $190$           & $74$            & $145$           & $81$            \\
\noalign{\smallskip}\hline
\end{tabular}\\
The values of $c_0$ for ELODIE and CORALIE are given relatively to their first measuments.
The uncertainties of the estimations are given in the brackets (e.g., $0.30(10)$ means
$0.30\pm 0.10$, and $30.0(1.0)$ means $30.0\pm 1.0$). The values for the mean longitudes
$\lambda$ and time shift parameters $\tau$ are given for the epoch $JD2452000$. The
uncertainties of the minimum masses $m\sin i$ and of the semi-major axes $a$ incorporate
the $10\%$ uncertainty of the stellar mass. The estimations of the effective RV jitters
$\sigma_\star^2$ incorporate an analytic correction of the statistical bias as discussed
in \citep{Baluev08b}.
\label{tab_bestfits}
\end{table}

Fig.~\ref{fig_Pc-Pd} shows such plot in the plane of orbital periods $P_c$ and $P_d$. We
can see that the likelihood function constructed from the full available dataset (top
panels of Fig.~\ref{fig_Pc-Pd}) has two main maxima with comparable values of $\tilde l$.
The first one is centred on $P_c\approx 2000$~days and $P_d\approx 870$~days, and the
second one on $P_c\approx 1800$~days and $P_d\approx 900$~days. We can see that the latter
solution is close to the 2/1 MMR of the outer planets. Such orbital configurations are
remarkable because only low-order MMRs can prevent planets on high-eccentricity orbits
from close approaches and hence can make the whole system stable. This `double-headed'
shape of the likelihood function looks rather stable with respect to the choice of the RV
model. From now on, we consider mainly the two mentioned families of solutions: the first
one corresponds to the resonance 2/1 between planets `c' and `d', the second one is
outside of this resonance (but may cover some other MMRs of relatively low, e.g. 7/3 and
5/2). Hereafter, we use the notation `A' for the first family and `B' for the second one.

The full sets of estimated parameters for the three RV models are shown in
Table~\ref{tab_bestfits}. The respective minimum values of $\tilde l$ depend on the model
adopted. For the models I and III, the `B' solution provides formally better fit to the RV
data in comparison with the `A' one, but for the model~II the corresponding values of
$\tilde l$ are similar. Either model~II or model~III provide some improvement in the
goodness-of-fit $\tilde l$, with respect to the model~I.

Let us note that the similar structure of the likelihood function can be also seen in the
graphs constructed in the similar way for the RV model~I, but with the use of the Keck
data only (left-bottom panel in Fig.~\ref{fig_Pc-Pd}). However, the shape of the
likelihood surface is more complicated in this case: the broad (much broader than in the
top pictures) peak corresponding to the B-family is actually `double-headed' itself (i.e.,
it consists of two close peaks having $P_d\approx 850$~days and $P_d\approx 820$~days).
Moreover, the shape of the likelihood function constructed with the use of the Keck data
only is severely dependent on the adopted RV model. For the model~II, the B-family peak is
clearly split into two distinct and very distorted peaks (one with $P_c\approx 1900$~days,
$P_d\approx 870$, and another one with $P_c \approx 2200$~days and $P_d$ less than
$800$~days, see middle-bottom panel in Fig.~\ref{fig_Pc-Pd}). For the model~III, all the
former peak are merged into a single very broad peak (formally centred not far from the
the 2/1 MMR), which is continued to infinite $P_c$ (right-bottom panel in
Fig.~\ref{fig_Pc-Pd}). Such behaviour, the further splitting of the local likelihood
maxima and their severe sensitivity to the RV model, indicates that the analysis of the
Keck data alone would yield significantly less reliable results than the joint analysis of
all available data.

However, even with the use of the full RV dataset, almost all of the best fits possess
large values of the condition number $\mathcal C$, especially for the `A' solution. This
means that we should treat our results with a more care. In fact, no one of the best (in
the sense of the goodness-of-fit measure) fits from Table~\ref{tab_bestfits} can be
accepted. The values of the eccentricity $e_c$ (and often those of the $e_d$ as well) are
large and lead to a very soon disintegration of the corresponding orbital configurations.
For comparison, Table~\ref{tab_circfits} contains the estimations of parameters for the
system with the eccentricity $e_c$ or both the eccentricities $e_c,e_d$ fixed at zero.
These fits have much smaller values of $\mathcal C$, though worse goodness-of-fit.
Numerical integration showed their dynamical stability and regular evolution on the time
scales of (at least) $10^6$~yr, except for the fit II$'$A which showed some signs of
chaoticity at the time scale of $\sim 10^5$~yr, evidently due to a large $e_d=0.337$. It
is important that only the `A' group of solutions contains best-fitting circular orbits:
the best-fitting solutions from the `B' group approach the 2/1 MMR and softly turn into
`A' group of solutions when $e_c$ decreases.

\begin{table}
\caption{Low-eccentricity orbital solutions for the planetary system around HD~37124.}
\begin{tabular}{@{}lllllll@{}}
\hline\noalign{\smallskip}
parameter            & I$'$A, 2/1      & I$''$A, 2/1     & II$'$A, 2/1     & II$''$A, 2/1    & III$'$A, 2/1    & III$''$A, 2/1   \\
\noalign{\smallskip}\hline\noalign{\smallskip}
\multicolumn{7}{c}{planet b}\\
$P$~[days]           & $154.34(12)$    & $154.31(12)$    & $154.52(12)$    & $154.36(13)$    & $154.35(11)$    & $154.34(11)$    \\
$\tilde K$~[m/s]     & $28.81(97)$     & $28.27(96)$     & $27.60(92)$     & $29.32(97)$     & $27.96(83)$     & $28.20(82)$     \\
$\lambda$~[$^o$]     & $117.4(2.0)$    & $117.1(1.9)$    & $117.0(1.8)$    & $117.0(1.9)$    & $118.6(1.7)$    & $118.5(1.7)$    \\
$e$                  & $0.078(32)$     & $0.080(32)$     & $0.090(28)$     & $0.084(31)$     & $0.062(29)$     & $0.065(29)$     \\
$\omega$~[$^o$]      & $132(23)$       & $131(23)$       & $116(19)$       & $124(22)$       & $146(25)$       & $147(24)$       \\
$m\sin i$~[$M_{Jup}$]& $0.644(48)$     & $0.654(48)$     & $0.617(46)$     & $0.656(48)$     & $0.625(46)$     & $0.631(46)$     \\
$a$~[AU]             & $0.518(17)5$    & $0.518(17)4$    & $0.518(17)9$    & $0.518(17)5$    & $0.518(17)5$    & $0.518(17)5$    \\
\multicolumn{7}{c}{planet d}\\
$P$~[days]           & $908(11).0$     & $905(11).8$     & $899.3(6.9)$    & $904(12).1$     & $883(10).7$     & $881.9(8.5)$    \\
$\tilde K$~[m/s]     & $15.2(1.1)$     & $14.17(92)$     & $17.1(1.2)$     & $14.13(94)$     & $15.65(88)$     & $15.21(81)$     \\
$\lambda$~[$^o$]     & $322.3(4.1)$    & $324.3(4.4)$    & $322.9(3.2)$    & $324.3(4.3)$    & $328.9(3.6)$    & $329.8(3.6)$    \\
$e$                  & $0.174(88)$     & $0($fixed$)$    & $0.337(71)$     & $0($fixed$)$    & $0.095(71)$     & $0($fixed$)$    \\
$\omega$~[$^o$]      & $349(21)$       &    -            & $1.7(9.8)$      &    -            & $358(33)$       &     -           \\
$m\sin i$~[$M_{Jup}$]& $0.614(62)$     & $0.572(53)$     & $0.689(66)$     & $0.570(54)$     & $0.626(54)$     & $0.608(52)$     \\
$a$~[AU]             & $1.690(58)1$    & $1.687(58)4$    & $1.679(57)3$    & $1.685(58)2$    & $1.659(57)8$    & $1.657(56)6$    \\
\multicolumn{7}{c}{planet c}\\
$P$~[days]           & $1810(41).3$    & $1839(50).0$    & $1759(28).0$    & $1834(54).7$    & $1815(40).0$    & $1822(44).7$    \\
$\tilde K$~[m/s]     & $13.4(1.3)$     & $12.0(1.1)$     & $15.4(1.5)$     & $11.9(1.0)$     & $12.5(1.1)$     & $11.86(88)$     \\
$\lambda$~[$^o$]     & $310.7(4.7)$    & $308.5(5.2)$    & $318.8(3.7)$    & $310.5(5.2)$    & $303.0(5.1)$    & $301.8(4.7)$    \\
$m\sin i$~[$M_{Jup}$]& $0.682(81)$     & $0.611(67)$     & $0.774(89)$     & $0.607(67)$     & $0.638(68)$     & $0.604(60)$     \\
$a$~[AU]             & $2.678(98)0$    & $2.70(10)61$    & $2.627(92)3$    & $2.70(10)19$    & $2.682(98)5$    & $2.690(99)1$    \\
\noalign{\smallskip}\hline\noalign{\smallskip}
$c_1$~[$\frac{\rm m}{\rm s\cdot yr}$]%
                     &     -           &     -           &     -           &    -            & $1.18(26)$      & $1.22(24)$      \\
\noalign{\smallskip}\hline\noalign{\smallskip}
\multicolumn{7}{c}{ELODIE dataset}\\
$c_0$~[m/s]          & $64.2(4.3)$     & $64.9(4.3)$     & $63.9(4.5)$     & $65.0(4.4)$     & $68.4(4.5)$     & $69.1(4.3)$     \\
$A$~[m/s]            & $18.9(3.1)$     & $19.9(3.3)$     & $17.8(3.3)$     & $19.8(3.4)$     & $18.9(3.7)$     & $19.5(3.7)$     \\
$\tau$~[days]        & $167(18)$       & $164(17)$       & $167(20)$       & $164(17)$       & $158(17)$       & $156(16)$       \\
$\sigma_\star$~[m/s] & $6.8(2.2)$      & $7.1(2.2)$      & $7.6(2.2)$      & $7.4(2.2)$      & $6.7(2.3)$      & $6.7(2.2)$      \\
r.m.s.~[m/s]         & $11.95$         & $12.33$         & $12.53$         & $12.45$         & $12.08$         & $12.18$         \\
\multicolumn{7}{c}{CORALIE dataset}\\
$c_0$~[m/s]          & $5.0(4.2)$      & $4.5(4.0)$      & $5.1(4.5)$      & $4.2(4.0)$      & $6.5(4.5)$      & $6.3(4.3)$      \\
$\sigma_\star$~[m/s] & $11.6(4.0)$     & $10.5(3.8)$     & $12.8(4.2)$     & $10.3(3.9)$     & $12.7(4.1)$     & $12.1(4.0)$     \\
r.m.s.~[m/s]         & $17.42$         & $16.76$         & $17.68$         & $16.53$         & $17.67$         & $17.35$         \\
\multicolumn{7}{c}{Keck dataset}\\
$c_0$~[m/s]          & $7.56(75)$      & $7.89(76)$      & $9.52(93)$      & $8.9(1.0)$      & $7.88(61)$      & $8.04(61)$      \\
$A$~[m/s]            &     -           &     -           & $4.9(1.3)$      & $2.0(1.4)$      &     -           &     -           \\
$\tau$~[days]        &     -           &     -           & $134(13)$       & $103(34)$       &     -           &     -           \\
$\sigma_\star$~[m/s] & $3.36(55)$      & $3.51(56)$      & $2.21(55)$      & $3.41(55)$      & $2.15(56)$      & $2.21(57)$      \\
r.m.s.~[m/s]         & $4.218$         & $4.319$         & $3.436$         & $4.193$         & $3.394$         & $4.446$         \\
\noalign{\smallskip}\hline\noalign{\smallskip}
$d$                  & $18$            & $16$            & $20$            & $18$            & $19$            & $17$            \\
$\tilde l$~[m/s]     & $8.710$         & $8.776$         & $8.280$         & $8.769$         & $8.065$         & $8.057$         \\
$\mathcal C$         & $34$            & $29$            & $40$            & $32$            & $36$            & $31$            \\
\noalign{\smallskip}\hline
\end{tabular}\\
The same notes as in Table~\ref{tab_bestfits} to be applied here. In these fits, the
eccentricity $e_c$ or both the eccentricities $e_c$ and $e_d$ were fixed at zero. All
these configurations correspond to the 2/1 MMR. The estimations of the semi-major axes and
orbital periods are given with one or two excessive decimal digits (shown after two-digit
uncertainties enclosed in brackets) to allow an unambigious reproducing of the long-term
dynamics of these configurations (see Section~\ref{sec_model}). For example, $1810(41).3$
means $1810.3\pm 41$ and $2.70(10)19$ means $2.7019\pm 0.10$.
\label{tab_circfits}
\end{table}

From the interplay of the indicators $\mathcal C$ and $\tilde l$ described above, we
derive the following conclusion. Although the orbital fits from Table~\ref{tab_bestfits}
show relatively small scatter of the residuals, this small scattering is in fact
fictitious, as indicated by the corresponding values of $\mathcal C$. The number of RV
data points and their temporal coverage still cannot constrain the full set in the model
parameters reliably. The full RV model is `overloaded'. Injecting some kind of \emph{a
priori} information (e.g., fixing the eccentricities at low values) allows to overcome the
obstacle of the statistical ill-determinacy. The resulting orbital fits possess better
statistical reliability, but by the cost of some increase of the RV residuals scatter.
Nevertheless, this increase is necessary to obtain a physically realistic orbital
configuration.

However, we cannot rule out the possibility that the actual orbits of the planets `c',`d'
are far from circular. To find more realistic orbital configurations than those from
Table~\ref{tab_bestfits}, but corresponding to eccentric orbits, we need to account for
more subtle requirements of the dynamical stability in our analysis.

\section{Dynamical interpretation}
\label{sec_dynam}
To obtain more realistic stable orbital configurations for this planetary system, we
continue to use the method of planar plots of partially minimized goodness-of-fit
statistic $\tilde l$. But now we examine orbital solution from a two-dimensional grid more
carefully: for each solution, we perform a numerical integration in order to rule out
rapidly disintegrating configurations. To perform such integrations, we need to know true
masses of planets in the system. As it can be seen from~(\ref{mass-SMA}), they depend on
the mass of the star $M_\star$ and on the orbital inclinations. Following \citet{Vogt05},
we adopt $M_\star = 0.78 M_\odot$ with an uncertainty of $10\%$. Unfortunately, using the
Keplerian RV model we can estimate only the minimum masses $m\sin i$ where the inclination
$i$ remains unknown. Until the gravitational interactions between planets in the system
are directly observed in the RV curve, the best thing that we can do is to assume \emph{a
priori} that the orbits are coplanar with $i=90^\circ$. If the actual inclination is less
than $90^\circ$, the true masses of the interacting planets become larger and the
stability region of the system become more narrow than for the edge-on configurations. The
same effect is expected from non-zero mutual inclinations of the orbits, due to the
well-known phenomenon of the $e$--$i$ coupling.

\begin{figure}
\begin{tabular}{@{}c@{}c@{}}
  \includegraphics[width=0.50\linewidth]{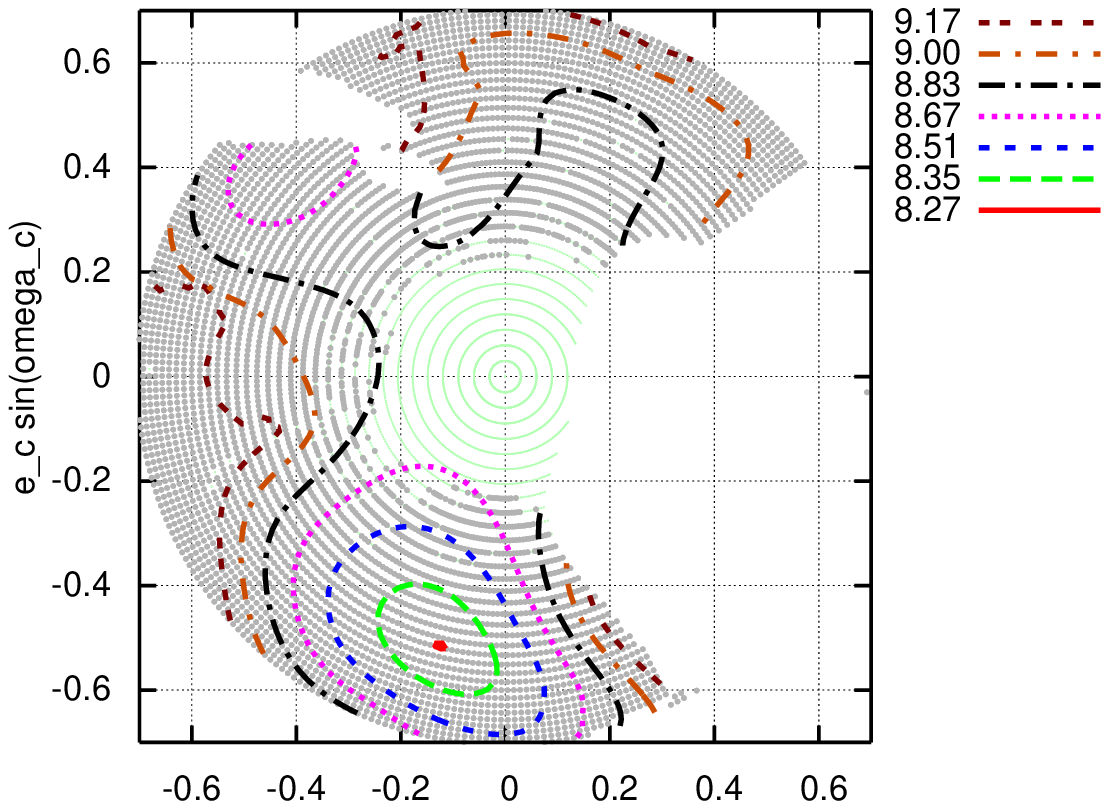} &
  \includegraphics[width=0.50\linewidth]{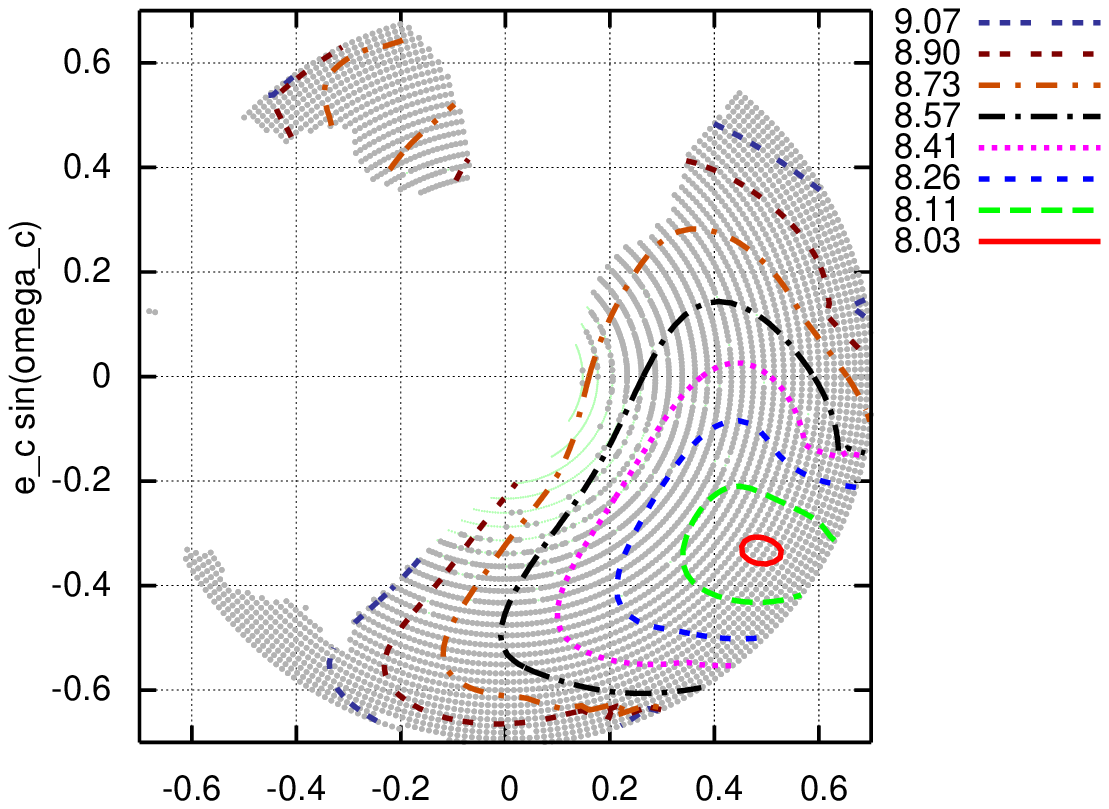} \\
  \includegraphics[width=0.50\linewidth]{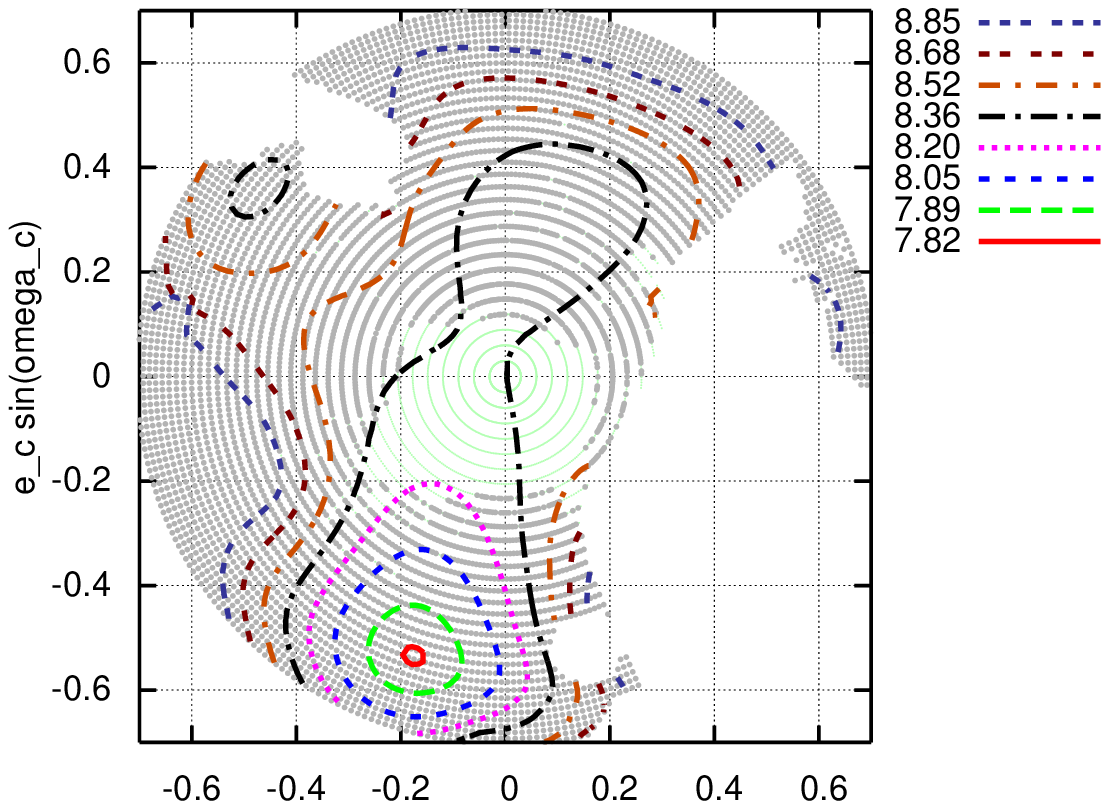} &
  \includegraphics[width=0.50\linewidth]{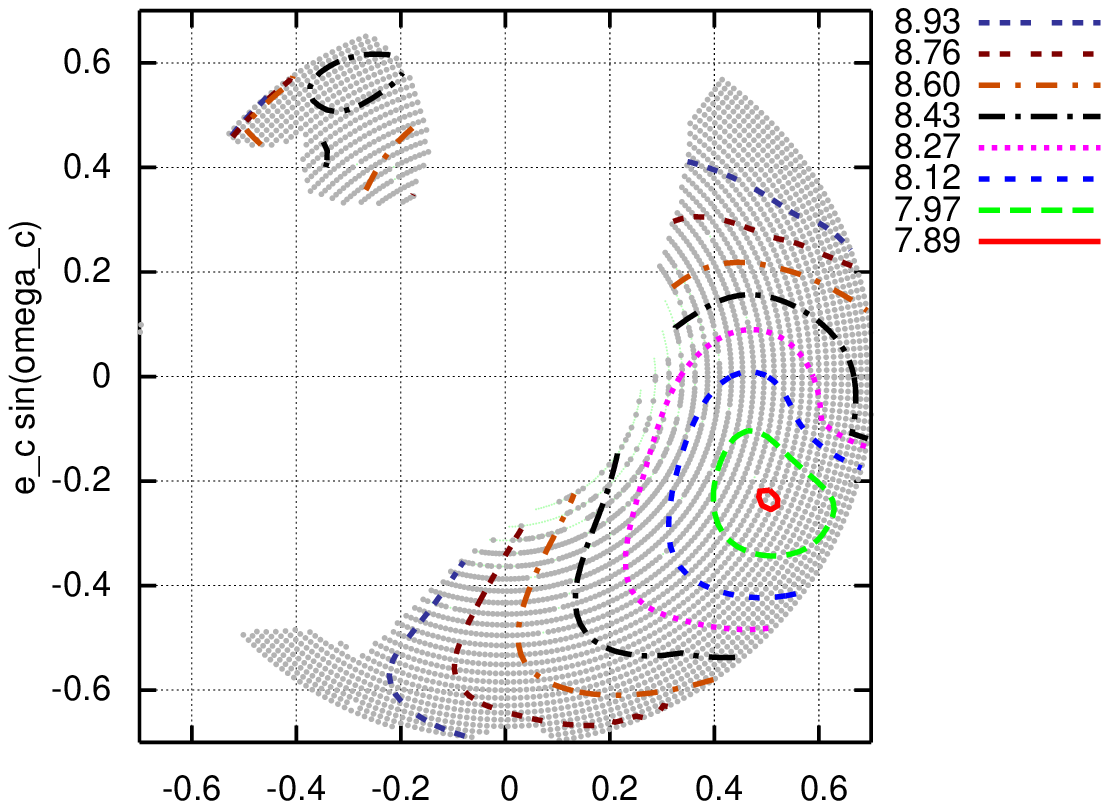} \\
  \includegraphics[width=0.50\linewidth]{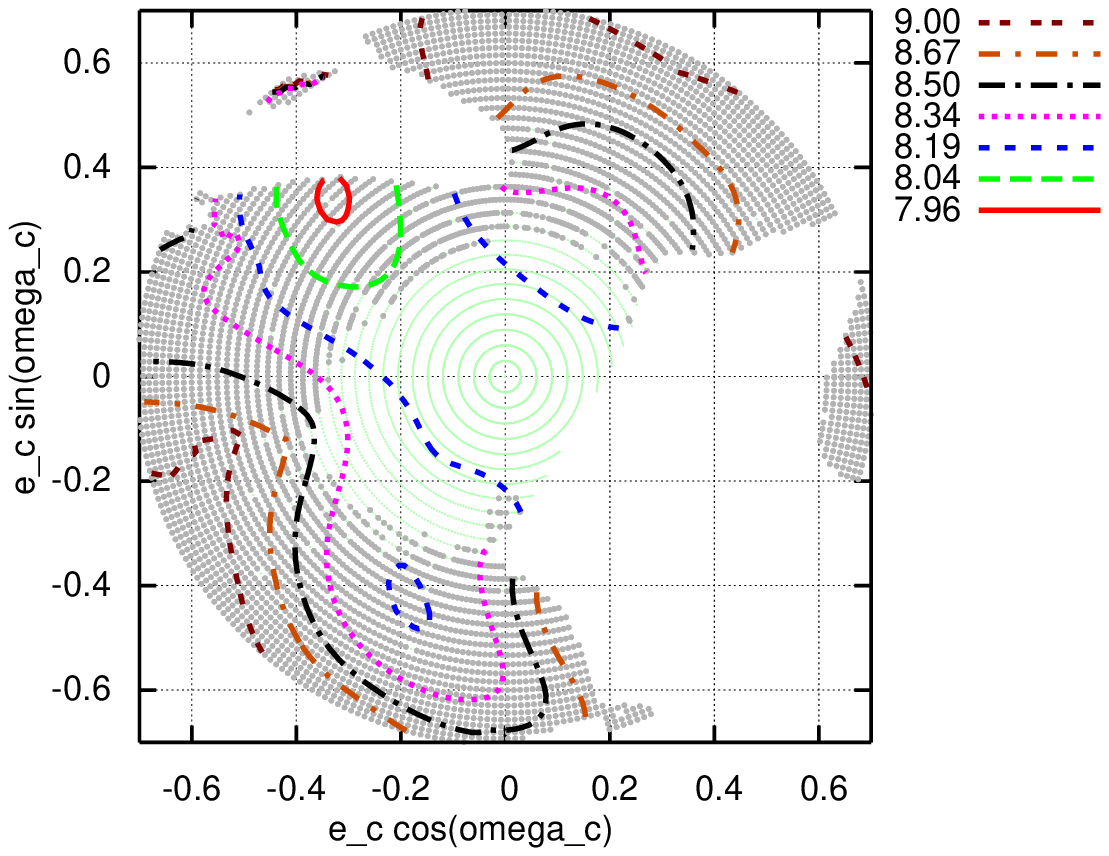} &
  \includegraphics[width=0.50\linewidth]{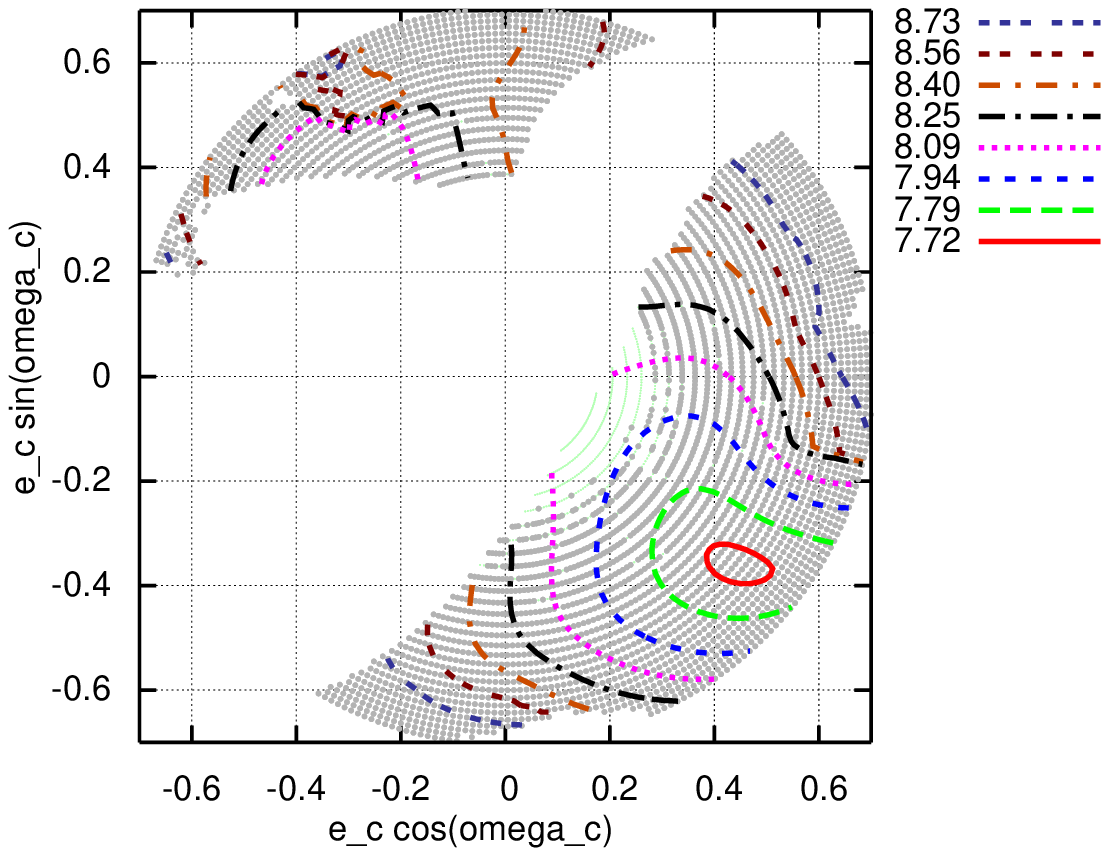} \\
\end{tabular}
\caption{Contour maps of the likelihood goodness-of-fit statistic for RV fits of HD37124.
These maps are plotted on the plane $(e_c \cos\omega_c, e_c \sin\omega_c)$ in the same way
as in Fig.~\ref{fig_Pc-Pd}. Solutions which correspond to orbital configurations
disintegrating in less than $30000$ years are marked by bold grey dots, other solutions
are marked by fine (green in the electronic version of the paper) dots. These dots are
arranged according to the polar coordinate system. Plots in the left column correspond to
the `A' solutions close to the 2/1 resonance (with no more than $5\%$ relative deviation
of the period ratio), plots in the right column correspond to the `B' solutions. The white
regions (where $e_c<0.7$) mark the points for which the fitting algorithm could not find a
solution from the corresponding family and switched to another one (that appeared
significantly more likely). The `A' and `B' families of solutions overlap in the region of
large $e_c$ without possibility of any smooth seam, but they seem to be sewed smoothly in
the region of small $e_c$. The top, middle, and bottom pairs of panels correspond to the
RV models I, II, and III, respectively.}
\label{fig_ecos-esin}
\end{figure}

Since much troubles in obtaining a realistic orbital configuration of the system are due
to the large eccentricity of the outermost planet, let us firstly consider the plane of
eccentric variables $(e_c\cos\omega_c,e_c\sin\omega_c)$. The corresponding maps are
plotted in Fig.~\ref{fig_ecos-esin}. We can see clearly the sophisticated shape of the
likelihood surface: among the `A' families of solutions, no one possess a single maximum.
Instead, we can see 2-3 local maxima; all of them correspond to large values of $e_c$. The
`B' family shows single maximum, but again at large $e_c$. No one of these local maxima
corresponds to a stable configuration. Stable solutions occupy only regions of small or
moderate $e_c$. It is important to note that the small-eccentricity solutions correspond
to the `A' configuration only: when $e_c$ decreases, a B-type solution approach the 2/1
resonance and softly turns into an A-type one.

As it can be seen from Fig.~\ref{fig_ecos-esin}, we can easily find stable solutions from
the `A' family. Such solutions can possess values of $\tilde l$ as small as $\approx
8.6$~m/s (model~I), $\approx 8.2$~m/s (model~II), and $\approx 8.1$~m/s (model~III). It is
interesting that the region of stable configurations is somewhat correlated with one of
the local minima of $\tilde l$ in the `A' layer, located in the third quadrant of the
coordinate plane. On contrary, we face much difficulties with obtaining a stable
configuration from the family `B'. The main reason for almost all `B' solutions to be
unstable is that the best-fitting period ratio $P_c/P_d\approx 2.1-2.3$ is quite small and
is not fixed (with a necessary precision) at any MMR of low order.

\begin{figure}\sidecaption
\begin{tabular}{@{}c@{}}
 \includegraphics[width=0.70\linewidth]{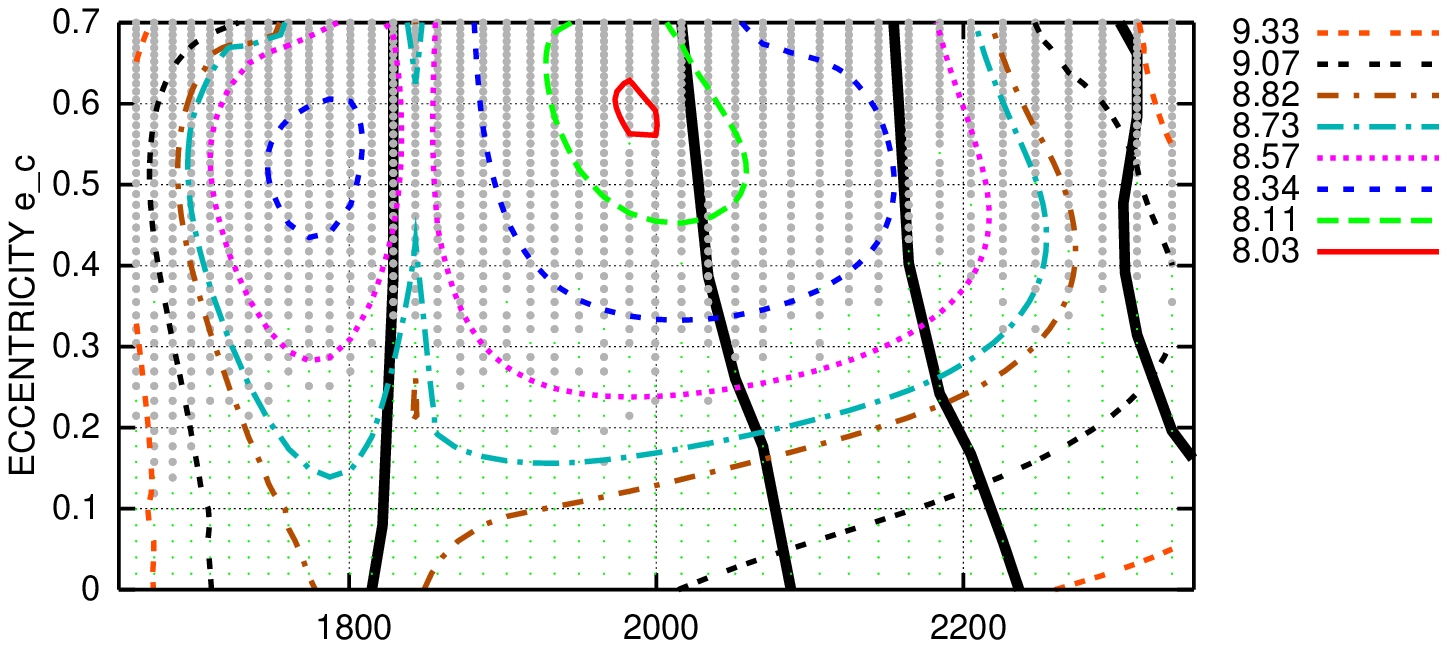}\\
 \includegraphics[width=0.70\linewidth]{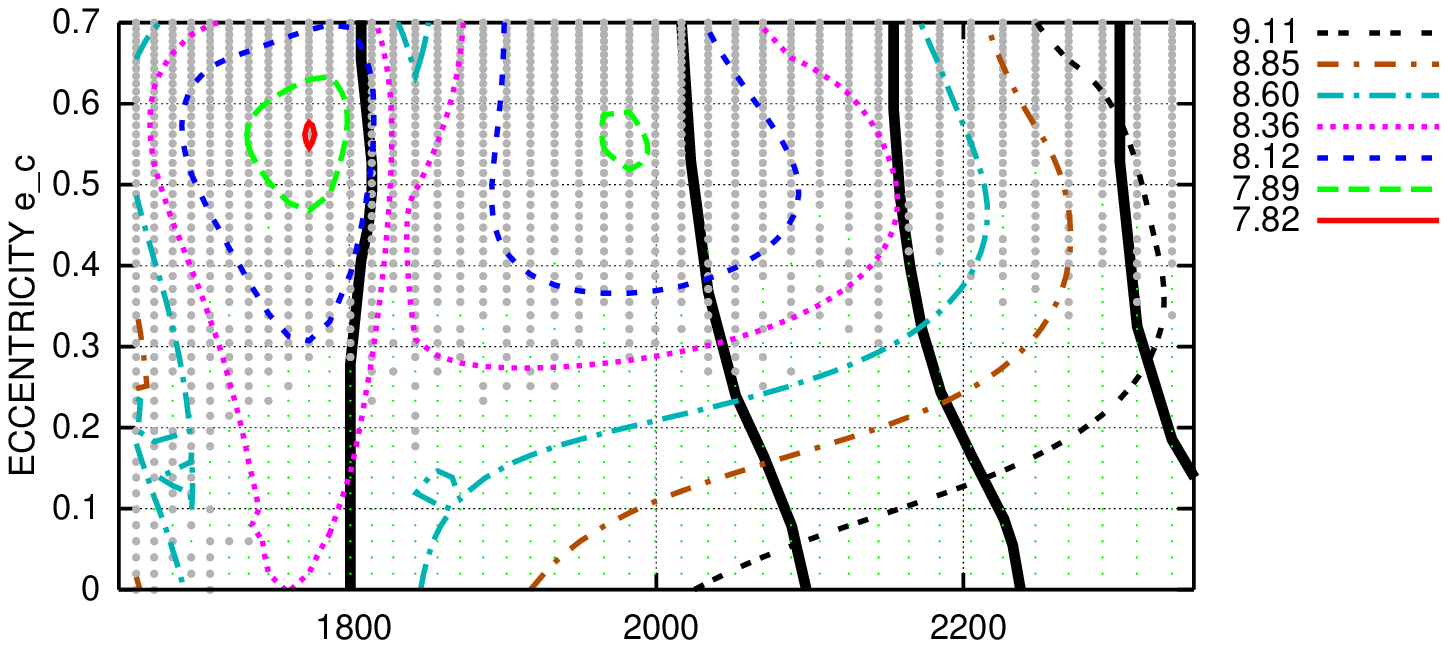}\\
 \includegraphics[width=0.70\linewidth]{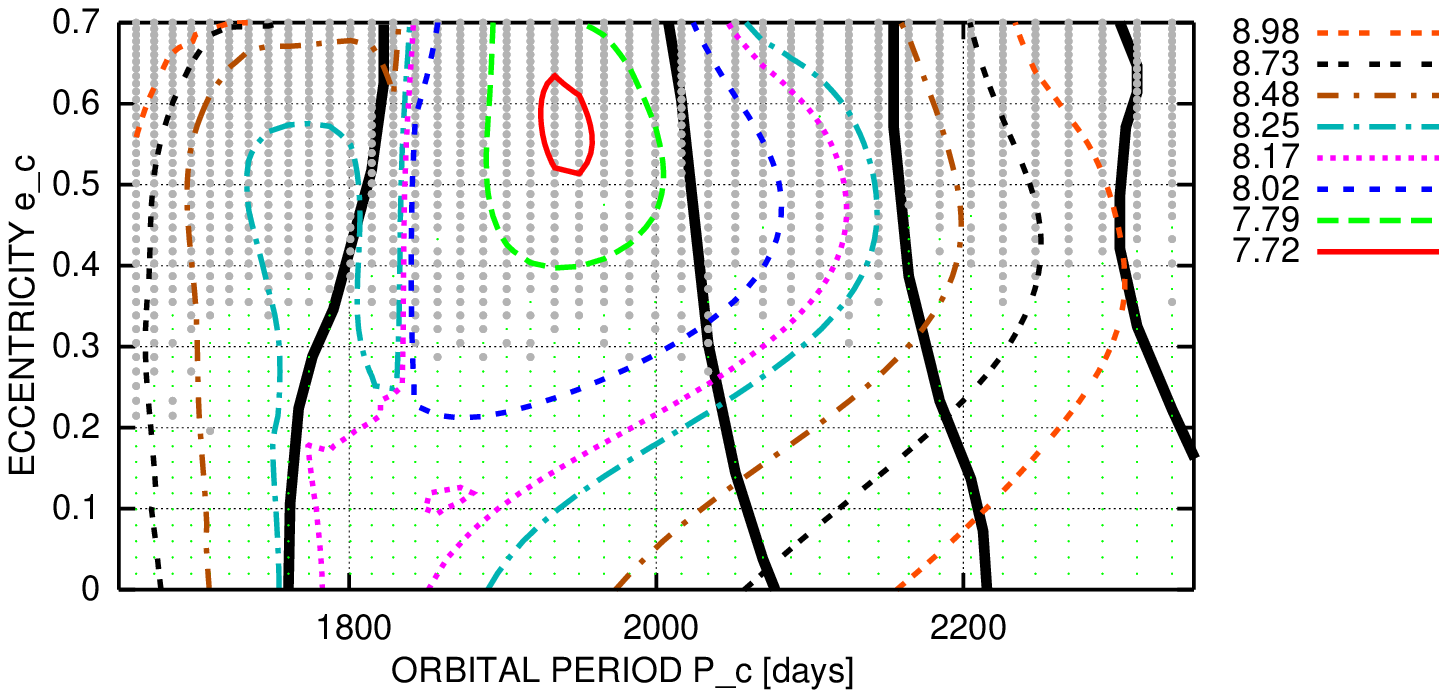}
\end{tabular}
\caption{Contour maps of the likelihood goodness-of-fit statistic for RV fits of HD37124.
These maps are plotted on the plane $(P_c, e_c)$ in the same way as in
Figs.~\ref{fig_Pc-Pd} and~\ref{fig_ecos-esin}. Three panels correspond to the models I,
II, and III (from top to bottom). In each of these panels, the A and B families of
solutions were merged in a single plot. The bold nearly vertical lines mark the solutions
having ratio of the best fitting periods $P_c$ and $P_d$ close to the 2/1, 7/3, 5/2, and
8/3 commensurabilities (lines from left to right).}
\label{fig_P-e}
\end{figure}

\begin{figure}
\begin{tabular}{@{}c@{}c@{}c@{}}
 \includegraphics[height=0.33\linewidth]{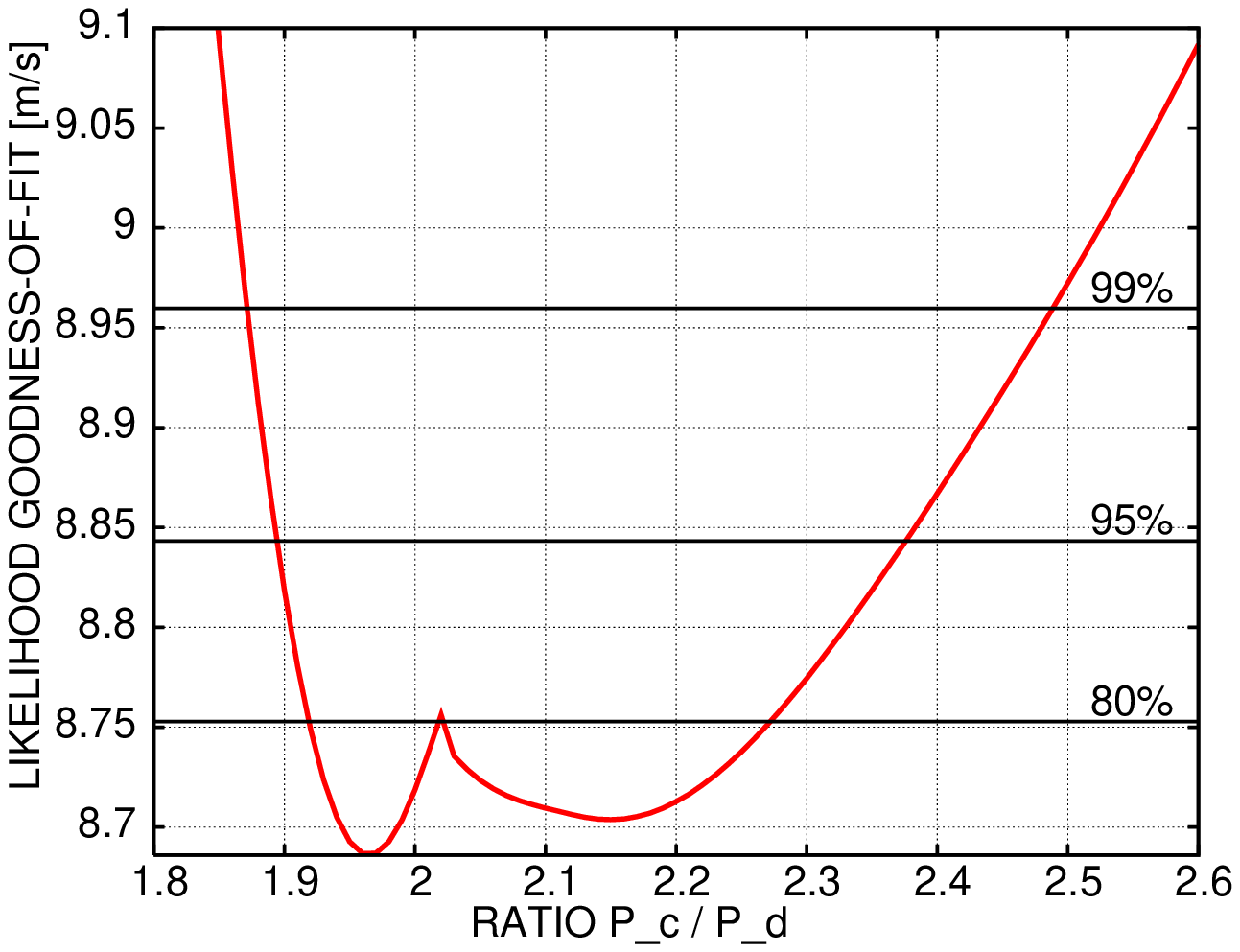} &
 \includegraphics[height=0.33\linewidth]{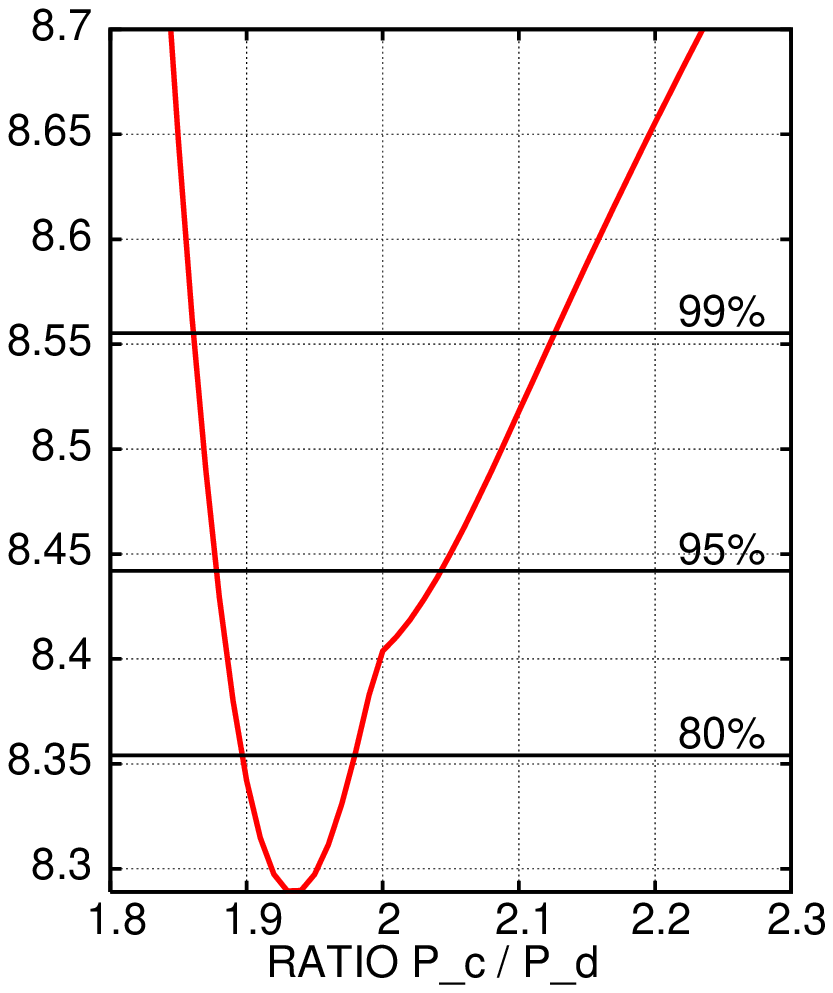} &
 \includegraphics[height=0.33\linewidth]{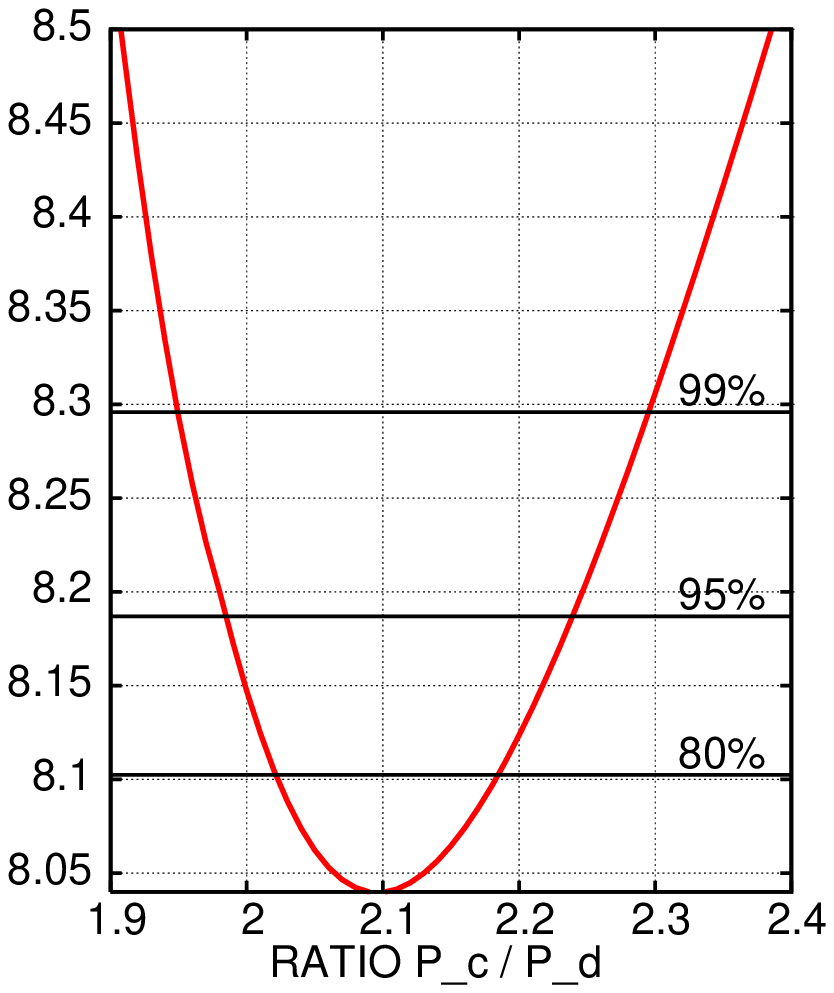}
\end{tabular}
\caption{Graphs of the likelihood goodness-of-fit function $\tilde l$, which was minimized
so that the eccentricity $e_c$ was fixed at $0.15$ and the ratio of orbital periods
$P_c/P_d$ was fixed at the values shown on the abscissas. Three panels correspond to the
RV model~I, II, and III (from left to right). The bold horizontal lines shows the levels
of $\min\tilde l$ yielding the $80\%$, $95\%$, and $99\%$ asymptotic confidence intervals
for $P_c/P_d$. These levels correspond to the values of the likelihood ratio statistic,
which provide the asymptotic false alarm probabilities as small as $20\%$, $5\%$, and
$1\%$
\citep[see][]{Baluev08b}.}
\label{fig_ratio}
\end{figure}

To locate stable solutions from the `B' group, we use another pair of variables. In
Fig.~\ref{fig_P-e}, the partially minimized $\tilde l$ is plotted in the plane $(P_c,
e_c)$. We can see that the fits with low $e_c$ and with $P_c$ fixed far from the 2/1
resonance, possess uncomfortably large values of $\tilde l$. Fig~\ref{fig_ratio}
illustrates this more clearly. In this figure, we plot the function $\tilde l$, minimized
so that the eccentricity $e_c$ was fixed at a safe value of $0.15$ and the period ratio
$P_c/P_d$ was fixed at the values marked on the abscissas. Since the value of the
eccentricity was fixed at a small value, the effects of the RV model non-linearity are
significantly suppressed (it follows from relatively small values of $\mathcal C$ in
Table~\ref{tab_circfits}), and thus we may try to find some confidence intervals for the
ratio $P_c/P_d$ using the standard asymptotic ($N\to\infty$) theory of point estimations.
We can see that the values $P_c/P_d>2.5$ and $P_c/P_d<1.87$ are beyond the $99\%$
confidence interval, when we use the RV model~I. For the other RV models, this confidence
interval becomes even smaller: $P_c/P_d\in [1.86,2.11]$ for the model~II and $P_c/P_d \in
[1.95,2.3]$ for the model~III. When the eccentricity $e_c$ decreases, these confidence
regions are shrinking around $P_c/P_d=2$.

However, we can note a promising region of \emph{high-eccentricity} solutions near the
resonance $5/2$ of the two outer planets. Surprisingly, numerical integration of the best
fitting configuration with $P_c$ fixed at $2170$~days and $e_c$ fixed at $0.4$ (in the RV
model~I) showed a quite regular evolution without any signs of instability at the time
scale of at least $10^6$ years. The evolution of this orbital configuration is a
large-amplitude oscillation around an antialigned apsidal corotation, Therefore, this
solution belongs to the class of orbital configurations in the 5/2 MMR found in
\citep{Gozd06} using only the Keck data.

It seems that, due to the non-linearity of the RV models coupled with lack of the data and
insufficient time coverage, none of the local minima of $\tilde l$ lies near the real
orbital configuration of this system. Probably, these multiple local minima are only the
fictitious `ripples' produced by the lack of the observations. All these `ripples' are
located deeply in the zone of dynamical instability, and therefore cannot be close to the
real configuration of the system. It is possible to find strictly the best-fitting orbital
solution(s) simultaneously satisfying the stability requirement. Evidently, such solutions
would be attracted by one of the `ripples', and therefore would be close to the boundary
of the domain of system stability. Therefore, we would have to deal with large
difficulties concerning the very complicated structure of the parameter space near such
boundaries \citep[see, e.g.,][]{Gozd08}, which probably represents the Arnold web
\citep[see, e.g,][]{Froeschle06}. Near such boundaries, the dynamics of the system is very
sensitive to small changes of parameters, the stability map of the parametric space is
strongly irregular, and hence the resolution of the parameter space should be chosen dense
enough. This requires very time-consuming calculations for checking the stability of probe
orbital configurations in this region. On contrary, it does not look likely that real
planetary systems can be found in such extremely dynamically active regions: it would be
rather difficult to explain how the system could migrate (without disintegration) to such
a state through the dense web of the instability threads and why it stopped in a thin
island of stability instead of moving further to dynamically unstable configurations.
Therefore, the reliability of such `hardly-stable' solutions would be too low to justify
the associated time-consuming calculations. In addition, considering the boundary of the
stability domain, it is rather difficult to understand the physical mechanism stabilizing
the configurations in the given domain.

In this paper, following \citet{Hadj06}, we will pay more attention to the \emph{centres}
of the stability domains which point out families of orbital configurations having some
`stock of stability'. For having a clear picture of possible dynamical regimes of a
planetary system, it is the position of the centre of the stability domain which should be
located and for which we should know possible uncertainties. Considering the kernel of a
stability domain, we avoid dealing with the sophisticated dynamical structure near its
boundary, which is hardly able to carry much information about the dynamics of the real
system. When an orbital solution has some stock of stability, the dynamics of the
corresponding planetary configuration is much more regular and much less sensitive to
small changes of parameters. To obtain such orbital solutions, we will try to decrease the
dimension of the problem (i.e., the number of degrees of freedom $d$) using certain
\emph{a priori} information about the stability of resonance planetary systems.

\section{The value of apsidal corotation resonances}
\label{sec_ACR}
Bearing in mind the results by \citet{Ji03,Hadj06,Hadj08,VoyatzisHadj06}, let us recall
that regular stable motions on high-eccentri\-city orbits with small period ratio are only
possible if the planets are trapped in a MMR and simultaneously are close to (or, at
least, not far from) an apsidal corotaion resonance. The details of the theory of apsidal
corotation resonances, along with necessary formulae and further references can be found,
for instance, in \citep{Beauge03}. For a brief summary, let us consider two planets
trapped in the $p/q$ MMR, i.e. having the ratio of orbital periods $P_2/P_1\approx p/q$
with $p>q$. We can write down the resonant angles
\begin{equation}
 s_1 = \frac{p \lambda_2 - q \lambda_1}{p-q} - \omega_1, \qquad
 s_2 = \frac{p \lambda_2 - q \lambda_1}{p-q} - \omega_2
\label{res-ang}
\end{equation}
and the canonically conjugated action variables
\begin{equation}
 I_1 = L_1 \left(1-\sqrt{1-e_1^2}\right), \qquad I_2 = L_2 \left(1-\sqrt{1-e_2^2}\right),
\label{conj-act}
\end{equation}
where $\lambda_i$ are the mean longitudes of the planets and $L_i\simeq m_i \sqrt a_i$ are
the Delaunay action variables. After averaging the Hamiltonian $H$ of the system over the
fast variables (i.e, over the mean longitudes $\lambda_i$) keeping the resting slow ones,
the resulting averaged Hamiltonian $\langle H \rangle$ depends on the canonical variables
$s_i,I_i$, and (as on parameters) on the masses $m_i$ of the planets. Evidently, this
averaging accounts properly for the orbital resonance. The averaged equations of motion
are then given by
\begin{equation}
 \frac{dI_i}{dt} = - \frac{\partial \langle H \rangle}{\partial s_i}, \qquad
 \frac{ds_i}{dt} =   \frac{\partial \langle H \rangle}{\partial I_i}
\label{hameq}
\end{equation}
Suppose that some values $s_i^*, I_i^*$ determine the position of an extremum of $\langle
H \rangle$. We can easily see that every such extremum provides a stationary solution
$s_i\equiv s_i^*, I_i\equiv I_i^*$ of the averaged system~(\ref{hameq}). Such stationary
solution is often called `apsidal corotation resonance' (hereafter ACR). If the initial
state of the planetary system slightly deviates from an exact ACR, the motion is a stable
oscillation around the exact stationary solution, because the planets are prevented from
close approaches. If the orbits of planets are highly eccentric and are far from
stationary solutions of the averaged Hamiltonian equations, the motion is, most probably,
highly chaotic and unstable: the secular drift of resonant angles~(\ref{res-ang}) leads
the planets to too close approaches destabilizing the system. Stable solutions with one or
both the resonant angles circulating are also possible in some cases; however, for
high-eccentricity configurations, the ACRs mark centres of dynamical stability \citep[see
e.g.][]{Hadj06,Hadj08}.

To obtain nominal orbital configurations of the system, we require from the resonant
planets `c' and `d' to be locked in an exact ACR (while neglecting the influence of the
innermost planet `b'). This can be justified not only by the stability considerations.
\citet{Beauge06} showed that adiabatic dissipative perturbations (e.g., interaction with
a protoplanetary disk) can cause planet pairs in a MMR to be captured in an ACR as well.

The requirement of the ACR lock implies four algebraic equations: $\partial\langle
H\rangle/\partial s_{c,d} = 0$ and $\partial\langle H\rangle/\partial I_{c,d} = 0 $. We
neglect here the gravitational influence of the innermost planet `b': it is seemingly
non-resonant with the outer planets and probably should not affect their resonant dynamics
much. The four equations mentioned above put certain constraints on the full set of free
parameters to be estimated from RV data and decrease the number of degrees of freedom by
four. It is very important, because this decreasing makes the problem significantly better
determined: we have about $8$ observations per a degree of freedom instead of about $6$.

\begin{table}
\caption{ACR fits for the planetary system around HD~37124.}
\begin{tabular}{@{}lllllll@{}}
\hline\noalign{\smallskip}
parameter            & I$^c$A, 2/1     & I$^c$B, 5/2     & II$^c$A, 2/1    & II$^c$B, 5/2    & III$^c$A, 2/1~   & III$^c$B, 5/2   \\
\noalign{\smallskip}\hline\noalign{\smallskip}
\multicolumn{7}{c}{planet b}\\
$P$~[days]           & $154.38(12)$    & $154.36(12)$    & $154.46(12)$    & $154.36(13)$    & $154.38(12)$    & $154.35(12)$    \\
$\tilde K$~[m/s]     & $29.21(92)$     & $28.9(1.0)$     & $28.84(83)$     & $28.9(1.0)$     & $28.69(85)$     & $28.8(1.0)$     \\
$\lambda$~[$^o$]     & $119.3(1.9)$    & $119.1(1.8)$    & $118.8(1.7)$    & $119.1(1.9)$    & $120.1(1.8)$    & $119.8(1.8)$    \\
$e$                  & $0.079(32)$     & $0.074(34)$     & $0.095(29)$     & $0.074(34)$     & $0.066(31)$     & $0.067(34)$     \\
$\omega$~[$^o$]      & $130(23)$       & $154(23)$       & $119(18)$       & $154(24)$       & $136(26)$       & $164(26)$       \\
$m\sin i$~[$M_{Jup}$]& $0.653(48)$     & $0.647(48)$     & $0.645(47)$     & $0.647(49)$     & $0.642(47)$     & $0.644(48)$     \\
$a$~[AU]             & $0.518(17)6$    & $0.518(17)5$    & $0.518(17)8$    & $0.518(17)5$    & $0.518(17)6$    & $0.518(17)5$    \\
\multicolumn{7}{c}{planet d}\\
$P$~[days]           & $907.6(8.9)$    & $874.6(7.8)$    & $897.2(7.8)$    & $874.4(8.2)$    & $896(11).3$     & $874.6(7.5)$    \\
$\tilde K$~[m/s]     & $16.10(81)$     & $13.82(86)$     & $16.53(81)$     & $13.84(91)$     & $16.63(78)$     & $14.40(91)$     \\
$\lambda$~[$^o$]     & $315.9(3.0)$    & $340.7(3.6)$    & $316.4(2.5)$    & $340.7(3.7)$    & $317.4(2.9)$    & $339.7(3.5)$    \\
$e$                  & $0.306(88)$     & $0.221(26)$     & $0.311(77)$     & $0.221(26)$     & $0.289(73)$     & $0.199(33)$     \\
$\omega$~[$^o$]      & $320.8(8.8)$    & $154.5(8.4)$    & $334.4(7.9)$    & $154.6(8.5)$    & $315(11)$       & $154.0(8.6)$    \\
$m\sin i$~[$M_{Jup}$]& $0.650(54)$     & $0.551(50)$     & $0.665(55)$     & $0.552(51)$     & $0.668(55)$     & $0.574(53)$     \\
$a$~[AU]             & $1.689(57)7$    & $1.648(56)3$    & $1.676(57)8$    & $1.648(56)1$    & $1.675(58)6$    & $1.648(56)4$    \\
\multicolumn{7}{c}{planet c}\\
$P$~[days]           & $1815(17).3$    & $2186(20).4$    & $1794(15).5$    & $2186(21).0$    & $1792(22).8$    & $2187(19).0$    \\
$\tilde K$~[m/s]     & $12.6(1.1)$     & $10.54(93)$     & $12.8(1.1)$     & $10.54(95)$     & $12.5(1.0)$     & $10.30(87)$     \\
$\lambda$~[$^o$]     & $311.8(5.3)$    & $301.0(5.5)$    & $318.2(5.0)$    & $301.0(5.6)$    & $307.7(6.0)$    & $300.3(5.5)$    \\
$e$                  & $0.122(56)$     & $0.377(74)$     & $0.132(42)$     & $0.379(78)$     & $0.136(54)$     & $0.333(76)$     \\
$\omega$~[$^o$]      & $267(23)$       & $334.5(8.4)$    & $278(21)$       & $334.6(8.5)$    & $250(20)$       & $334.0(8.6)$    \\
$m\sin i$~[$M_{Jup}$]& $0.639(70)$     & $0.570(63)$     & $0.648(71)$     & $0.570(64)$     & $0.631(68)$     & $0.558(60)$     \\
$a$~[AU]             & $2.683(91)0$    & $3.03(10)69$    & $2.662(90)4$    & $3.03(10)66$    & $2.660(91)8$    & $3.03(10)75$    \\
\noalign{\smallskip}\hline\noalign{\smallskip}
$c_1$~[$\frac{\rm m}{\rm s \cdot \rm yr}$]%
                     &     -           &    -            &     -           &     -           & $0.87(32)$      & $0.44(26)$      \\
\noalign{\smallskip}\hline\noalign{\smallskip}
\multicolumn{7}{c}{ELODIE dataset}\\
$c_0$~[m/s]          & $64.7(4.4)$     & $65.5(4.7)$     & $64.6(4.6)$     & $65.5(4.5)$     & $67.4(4.5)$     & $66.8(4.5)$     \\
$A$~[m/s]            & $18.8(3.1)$     & $20.2(3.2)$     & $18.1(3.3)$     & $20.2(3.3)$     & $18.4(3.4)$     & $19.8(3.3)$     \\
$\tau$~[days]        & $169(19)$       & $167(17)$       & $168(20)$       & $167(18)$       & $163(18)$       & $165(18)$       \\
$\sigma_\star$~[m/s] & $7.2(2.2)$      & $7.7(2.1)$      & $8.2(2.1)$      & $7.7(2.1)$      & $6.8(2.2)$      & $7.6(2.1)$      \\
r.m.s.~[m/s]         & $12.20$         & $12.60$         & $12.85$         & $12.51$         & $12.15$         & $12.33$         \\
\multicolumn{7}{c}{CORALIE dataset}\\
$c_0$~[m/s]          & $4.8(4.1)$      & $6.2(3.8)$      & $4.5(4.1)$      & $6.1(3.8)$      & $5.7(4.4)$      & $6.3(3.9)$      \\
$\sigma_\star$~[m/s] & $11.5(3.9)$     & $9.4(3.8)$      & $11.3(4.0)$     & $9.4(3.8)$      & $12.6(4.1)$     & $10.0(3.8)$     \\
r.m.s.~[m/s]         & $17.91$         & $15.91$         & $17.43$         & $15.82$         & $18.44$         & $16.06$         \\
\multicolumn{7}{c}{Keck dataset}\\
$c_0$~[m/s]          & $7.13(73)$      & $8.13(72)$      & $9.1(1.0)$      & $8.1(1.0)$      & $7.11(66)$      & $8.15(71)$      \\
$A$~[m/s]            &     -           &     -           & $3.9(1.3)$      & $0.1(1.1)$      &     -           &     -           \\
$\tau$~[days]        &     -           &     -           & $131(17)$       &     -           &     -           &     -           \\
$\sigma_\star$~[m/s] & $3.48(56)$      & $3.15(56)$      & $2.77(54)$      & $3.27(55)$      & $2.93(55)$      & $3.08(55)$      \\
r.m.s.~[m/s]         & $4.322$         & $4.108$         & $3.762$         & $4.148$         & $3.910$         & $4.001$         \\
\noalign{\smallskip}\hline\noalign{\smallskip}
$d$                  & \multicolumn{2}{c}{$16$}          & \multicolumn{2}{c}{$18$}          & \multicolumn{2}{c}{$17$}          \\
$\tilde l$~[m/s]     & $8.851$         & $8.616$         & $8.601$         & $8.693$         & $8.530$         & $8.549$         \\
$\mathcal C$         & $39$            & $56$            & $38$            & $64$            & $48$            & $79$            \\
\noalign{\smallskip}\hline
\end{tabular}\\
The same notes as in Table~\ref{tab_bestfits} to be applied here. These orbital elements
were obtained for the Jacobi coordinate system with appropriate masses assigned to the
reference barycentres. See Appendix~\ref{sec_fitACR} for more details concerning the
algorithm used to obtain these fits. The semi-major axes and some orbital periods are
given with excessive decimal digits, as in Table~\ref{tab_circfits}.
\label{tab_apscorfits}
\end{table}

More detailed description of the algorithm used to obtain such ACR fits, is given in
Appendix~\ref{sec_fitACR}. The resulting fits of ACR solutions are given in
Table~\ref{tab_apscorfits}. It is important that since the position of the ACR depends on
the planetary mass ratio only and is almost independent of the individual planet masses,
the solutions from Table~\ref{tab_apscorfits} are almost independent of the assumptions
about the orbital inclination of the system: $(m_c\sin i)/(m_d\sin i) = m_c/m_d$. We only
need to adjust the ratio of orbital periods by the quantity $\mathcal O(m/M_\star)$ (see
formula~(\ref{PcPd}) in Appendix~\ref{sec_fitACR}), but this adjustment have insignificant
effect on the RV fit quality (until we consider inclinations as small as a few degrees).
This invariance with respect to orbital inclination represents an extra advantage of the
use of the ACR fitting procedure.

\begin{figure}
\begin{tabular}{@{}c@{}c@{}}
 \includegraphics[width=0.50\linewidth]{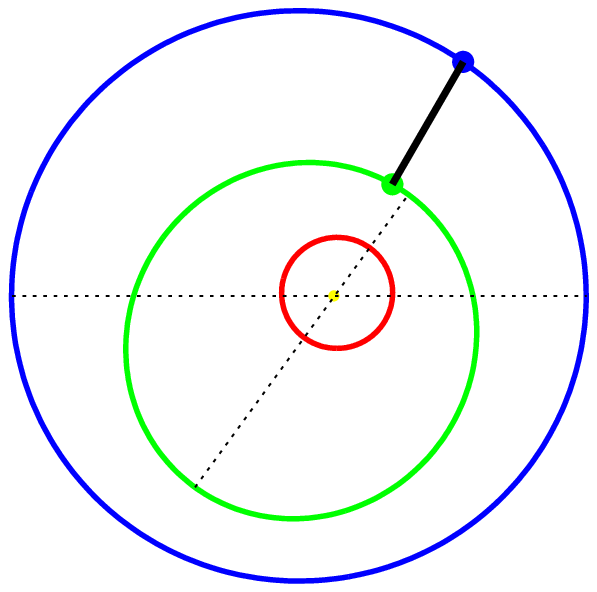} &
 \includegraphics[width=0.50\linewidth]{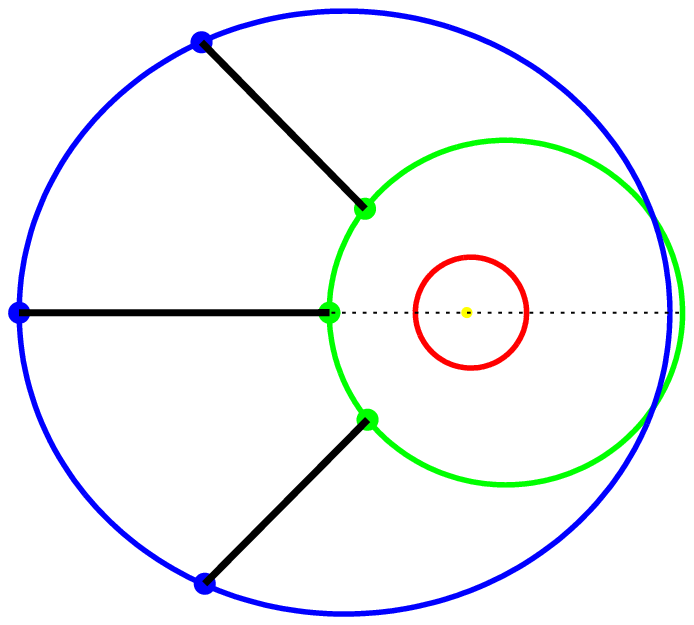} \\
 \includegraphics[width=0.50\linewidth]{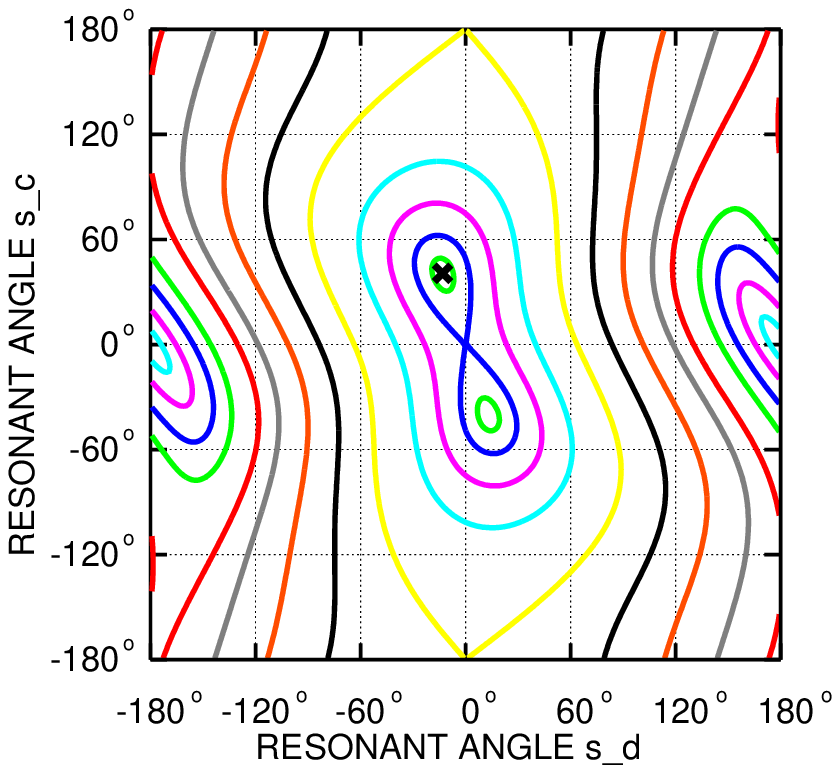} &
 \includegraphics[width=0.50\linewidth]{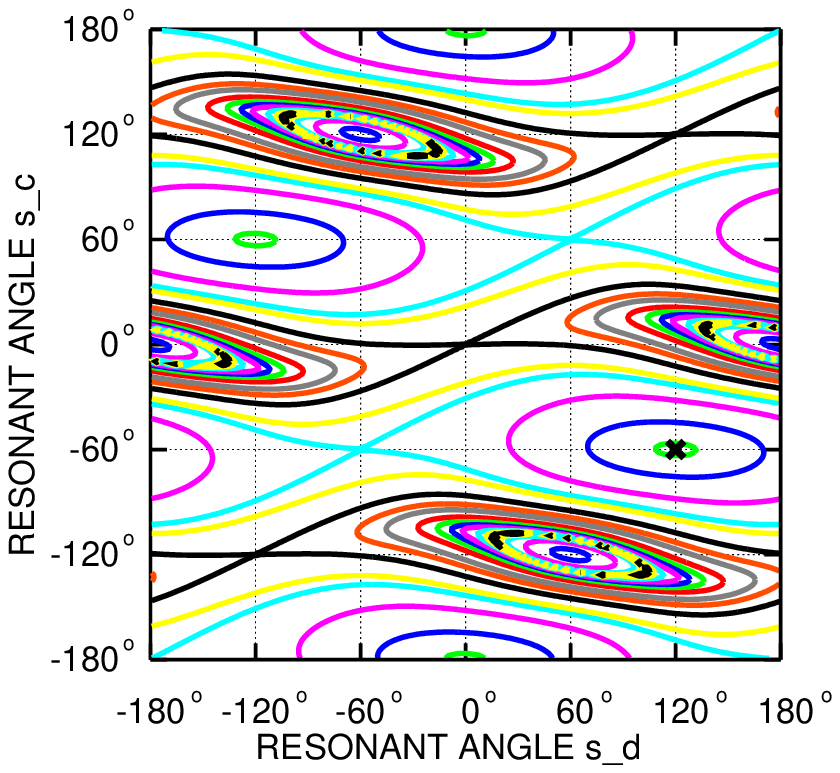}
\end{tabular}
\caption{Top panels: orbits of the HD37124 system for the ACR fits (model~I only). Left
panel is for the 2/1 MMR; right panel is for the 5/2 MMR. Straight solid lines mark the
conjunction positions. Broken lines mark the lines of apses. Middle panels: contour maps
of the averaged Hamiltonian $\langle H_{cd} \rangle$ in the planes of resonant angles
$s_d,s_c$, for the same MMRs and eccentricities as in top panels. Crosses mark the actual
ACR positions corresponding to local maxima of $\langle H_{cd}\rangle$.}
\label{fig_plan}
\end{figure}

\begin{figure}\sidecaption
\begin{tabular}{@{}c@{}}
 \includegraphics[width=0.70\linewidth]{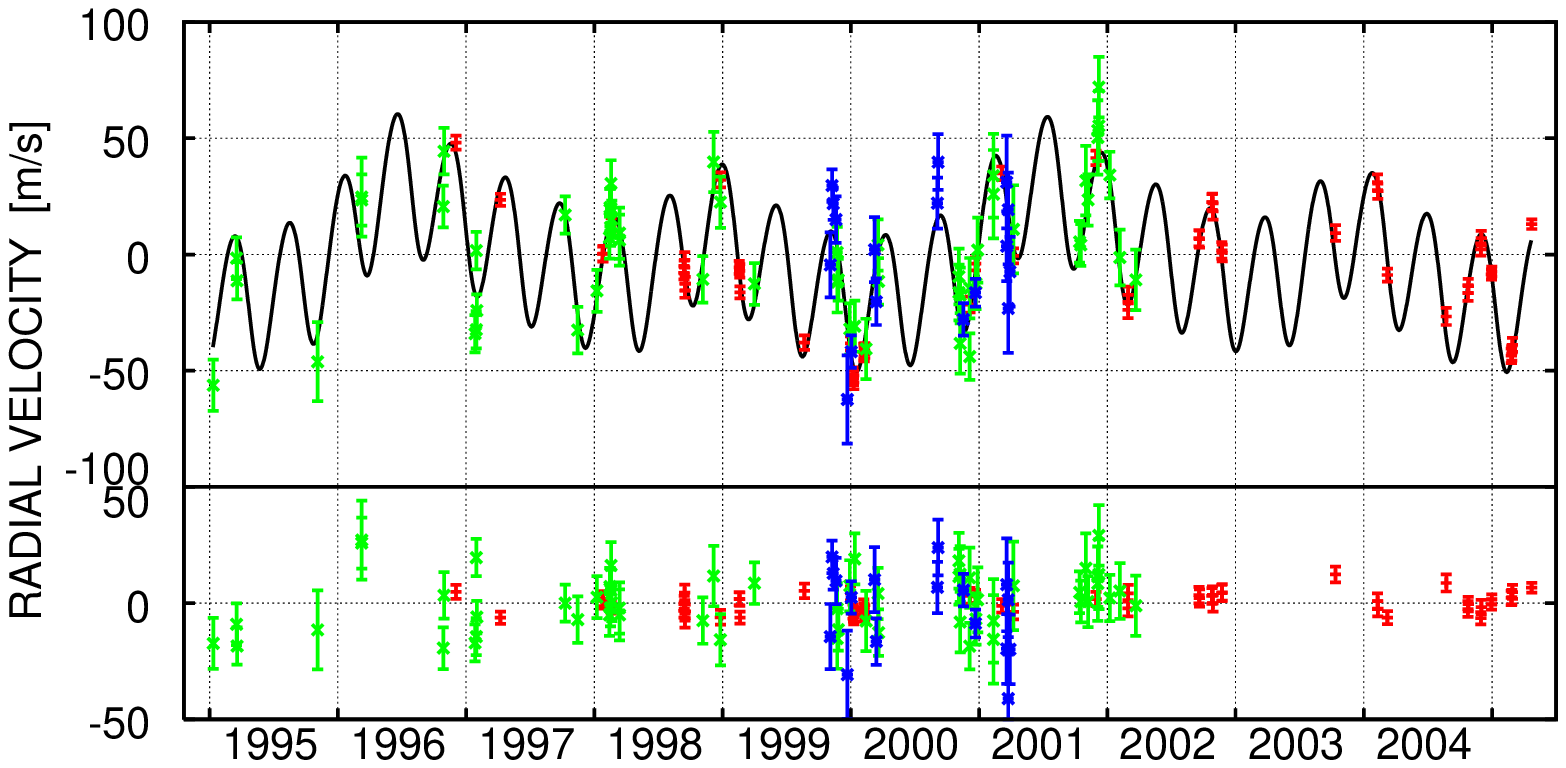}\\
 \includegraphics[width=0.70\linewidth]{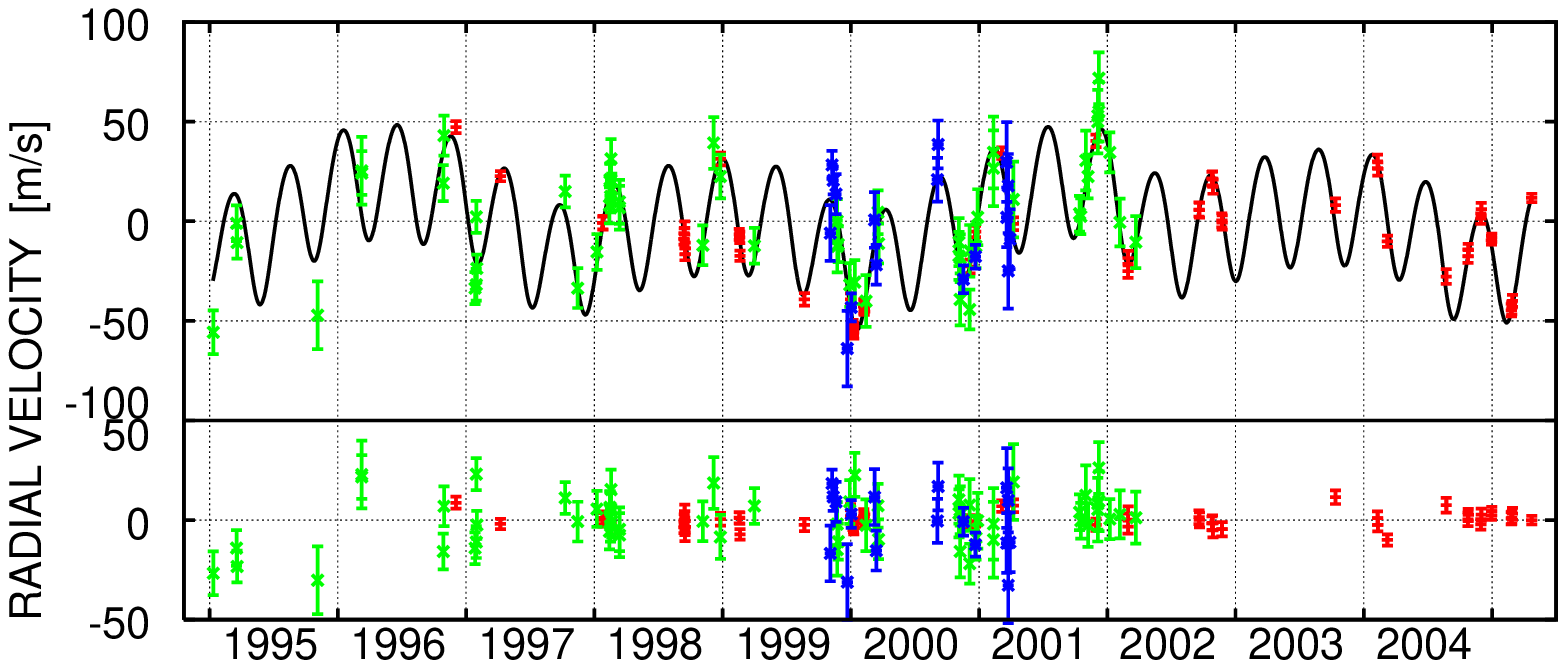}
\end{tabular}
\caption{RV curves for the ACR solutions I$^c$A (top panel) and I$^c$B (bottom panel),
plotted together with the RV measurements and their residuals. The error bars do not
incorporate estimated values of the RV jitter. The best-fitting annual sinusoidal drift of
the ELODIE measurements was preliminarily subtracted.}
\label{fig_RV-ACR}
\end{figure}

We can see that in the case of the 2/1 resonance, the best fitting ACR is asymmetric with
difference between the longitudes of periastra about $60^\circ$, whereas in the case of
the 5/2 resonance, the corresponding ACR is symmetric and antialigned (i.e.,
$\omega_c-\omega_d = 180^\circ$), see Fig.~\ref{fig_plan}. The corresponding RV curves fit
all available RV data satisfactorily, including the ELODIE measurements
(Fig.~\ref{fig_RV-ACR}). The values of the goodness-of-fit measure $\tilde l$ are not
increased very much with respect to those from Table~\ref{tab_bestfits} and are comparable
with those from Table~\ref{tab_circfits}. This means that one of the ACR fits can reflect
the true configuration of this system quite well. The corresponding condition numbers
$\mathcal C$ for the 2/1 resonance in Table~\ref{tab_apscorfits} are much less than in
Table~\ref{tab_bestfits}. This indicates that the topology of the likelihood surface
becomes simpler and closer to the desirable paraboloidal one. The troubles connected with
multiple local maxima of the likelihood have been overcome in the ACR fits. We can now say
more definitely, that the effect of the putative annual term in the Keck RV data on the
fit quality is similar to the effect of the linear RV trend (which could be due to a
long-period planet or brown dwarf in the system). These extra terms are significant in the
fits for the 2/1 resonance (the `A' group of solutions). For these fits, the estimations
of false alarm probabilities, calculated from the corresponding likelihood ratios
according to \citet{Baluev08b}, are about $2\%$ and $0.3\%$ (for the annual term and for
the linear drift, respectively). The 5/2 resonance (B) solution does not require these
terms in the RV model. This dilemma can be solved by future observations only.

\section{Long-term dynamics}
\label{sec_evol}
Now we consider the dynamical regimes of our nominal orbital configurations more closely.
We are especially interested in how much the innermost planet can disturb the apsidal
corotation of the two outer planets. This effect may be split in two categories:
\begin{enumerate}
\item The planet `b' can inspire some extra oscillation of the outer planets `c', `d' near
their unperturbed ACR (`unperturbed' means obtained without taking into account the
influence of the planet `b').
\item The planet `b' can shift the position of the libration centre from the unperturbed ACR.
\end{enumerate}
Only the first effect may significantly affect the system stability, because only a
large-amplitude oscillation around the libration centre can significantly increase the
probability of close approaches of the resonant planets.

\begin{figure}
\begin{tabular}{@{}c@{}c@{}}
  \includegraphics[width=0.49\linewidth]{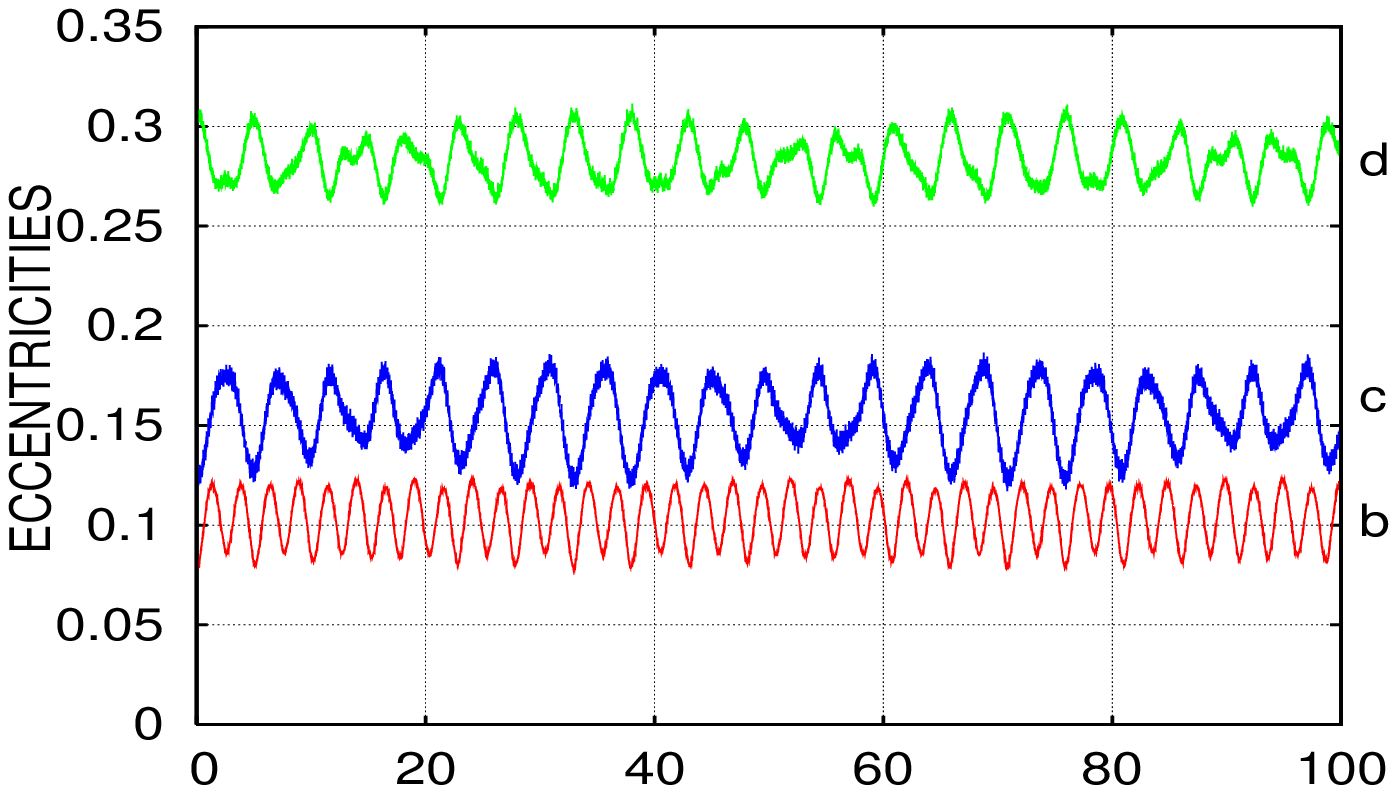} &
  \includegraphics[width=0.502\linewidth]{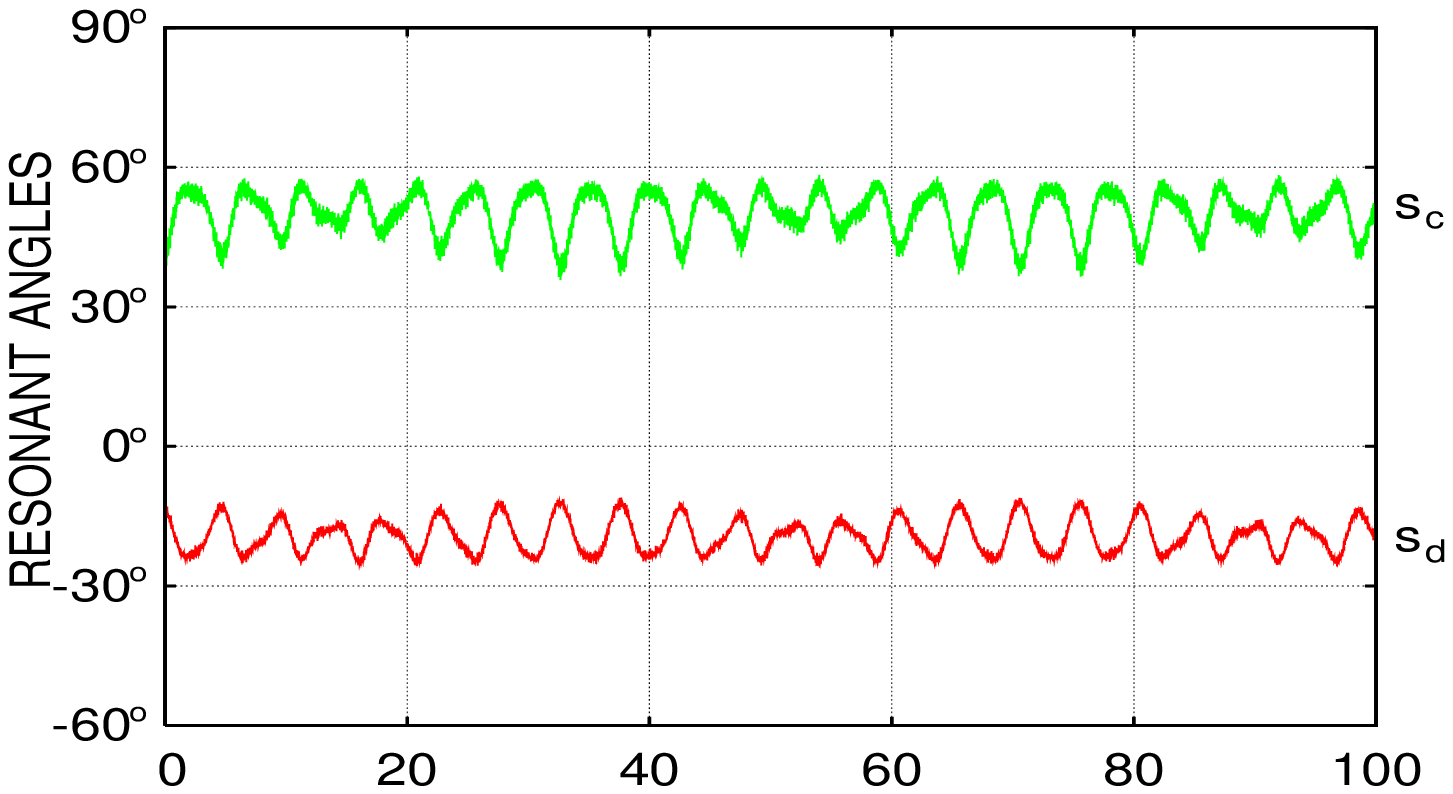} \\
  \includegraphics[width=0.49\linewidth]{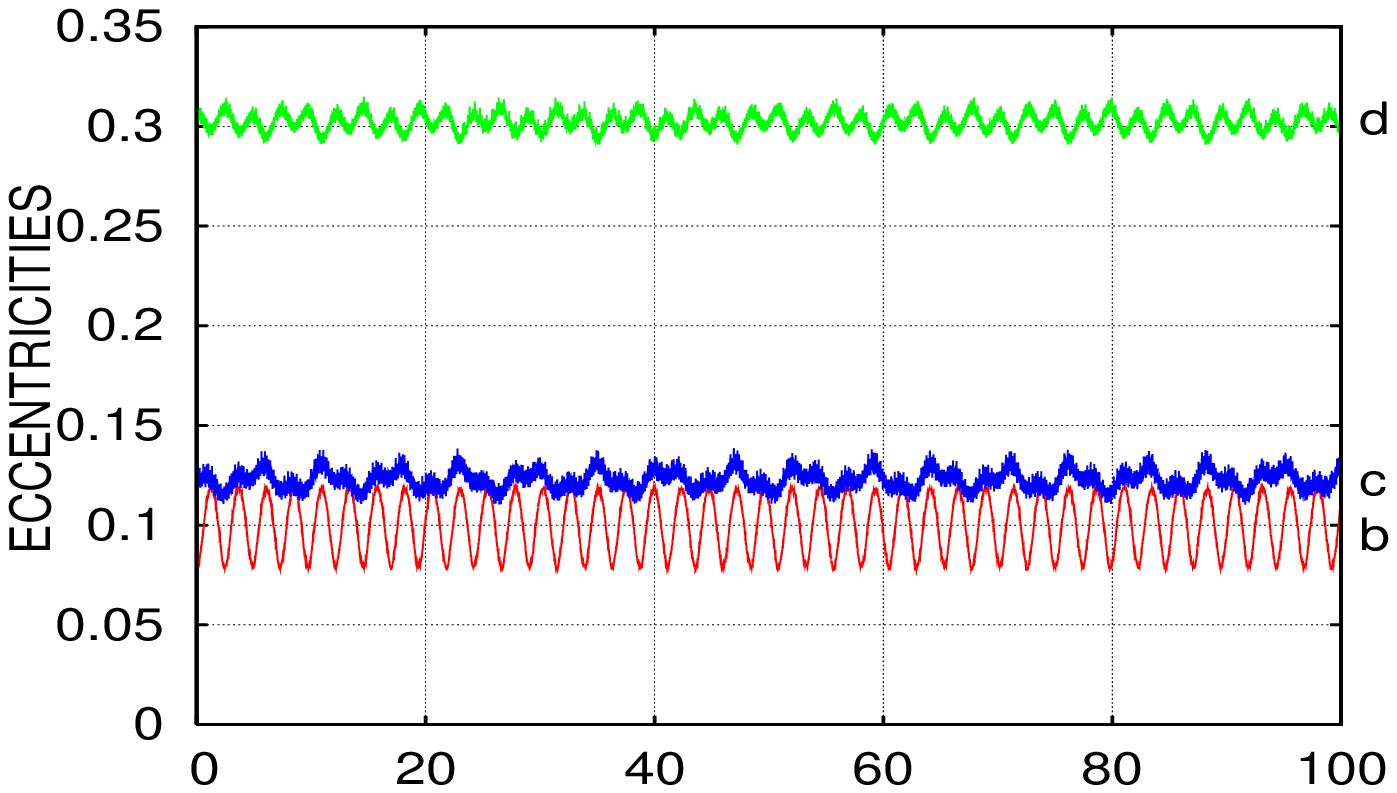} &
  \includegraphics[width=0.502\linewidth]{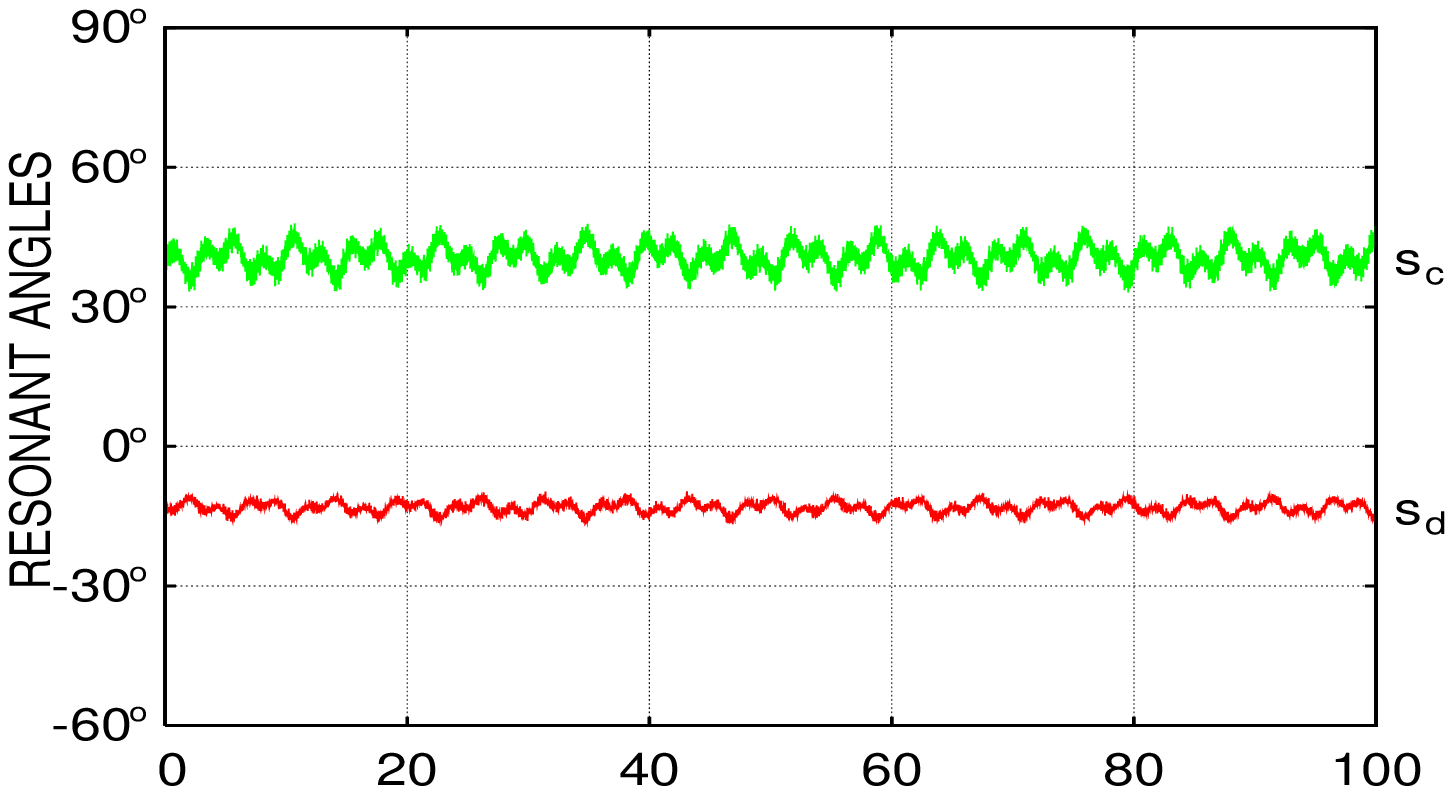} \\
  \includegraphics[width=0.49\linewidth]{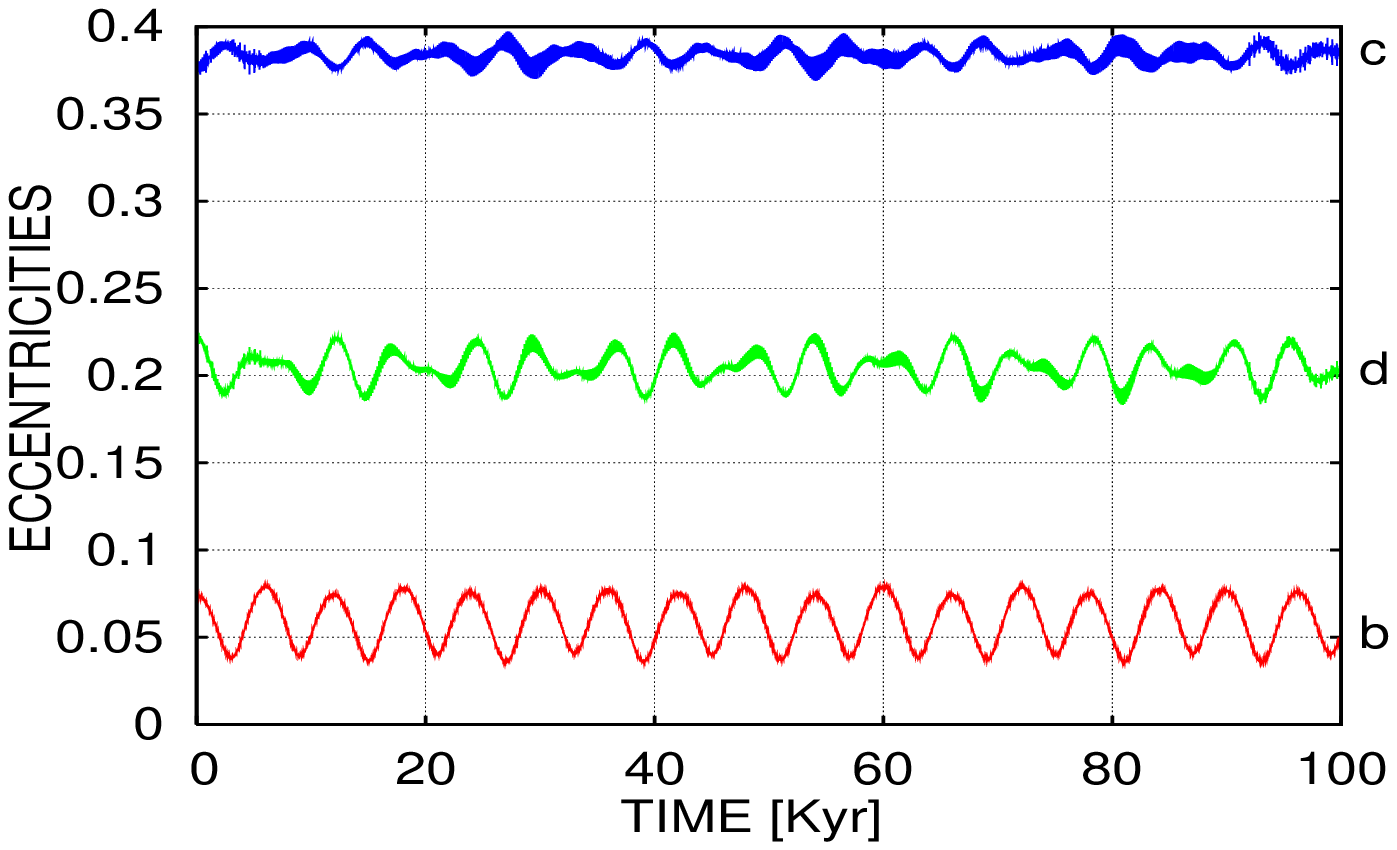} &
  \includegraphics[width=0.502\linewidth]{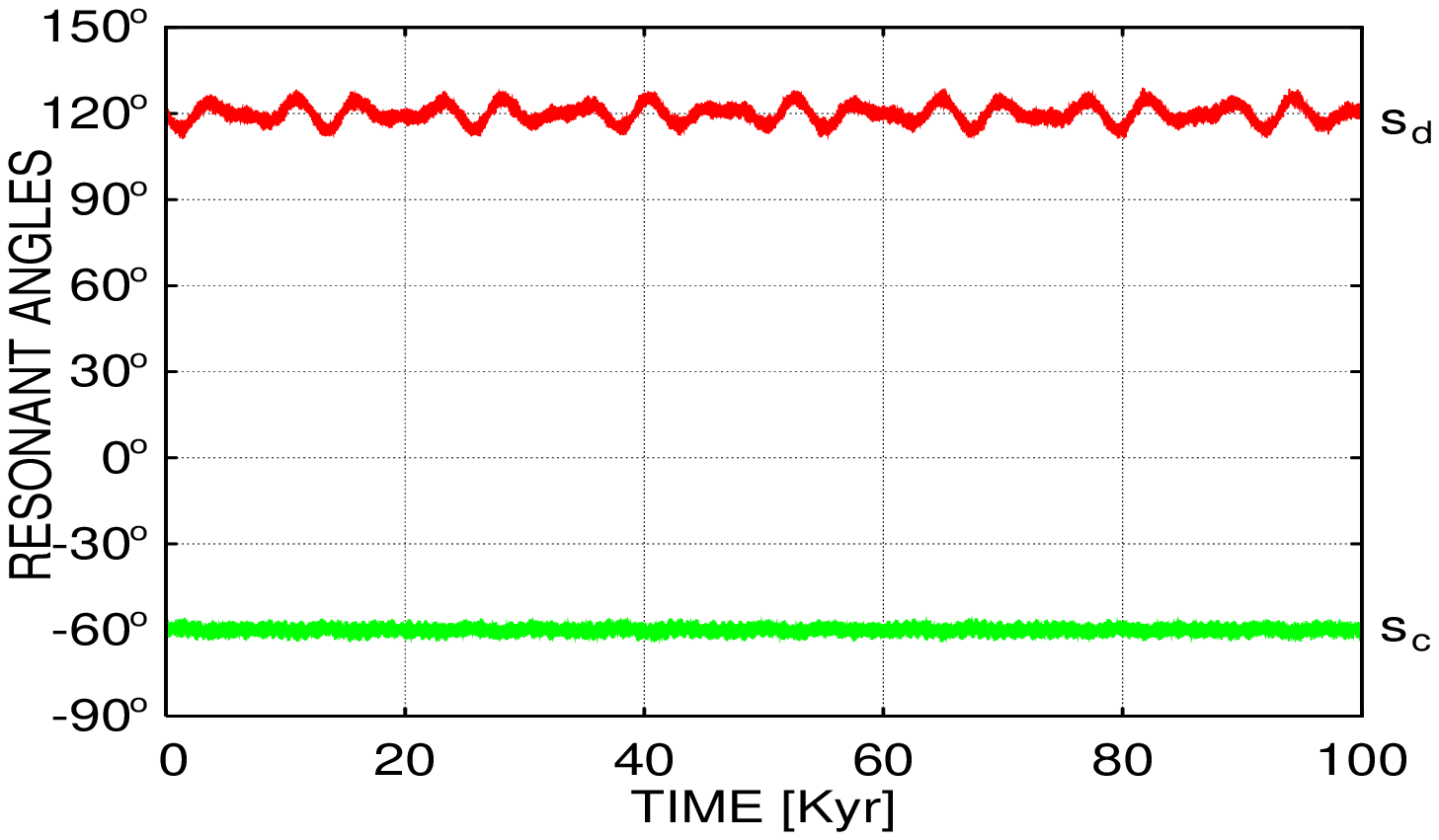} \\
\end{tabular}
\caption{Temporal evolution of the ACR configurations of planets in HD37124 (for RV model
I only). Left panels show the evolution of the eccentricities, right panels show the
evolution of resonant angles. Top pair: resonance 2/1, orbital parameters were taken from
Table~\ref{tab_apscorfits}, solution I$^c$A. Middle pair: the same case, but the value of
the mass $m_c$ was increased so that the value of $\tilde K_c$ increased by $1$~m/s.
Bottom pair: resonance 5/2 (Table~\ref{tab_apscorfits}, solution I$^c$B). The same
character of motion is conserved on the time scale of $10^6$~yr (and probably much longer)
for all the cases.}
\label{fig_evol}
\end{figure}

Fig.~\ref{fig_evol} illustrates both effects. We can see that the orbital configuration
I$^c$A taken from Table~\ref{tab_apscorfits} shows moderate oscillation of the
eccentricities $e_c,e_d$ and libration of resonant angles $s_c,s_d$ around some
equilibrium values. However, the centres of oscillations are somewhat shifted with respect
to the unperturbed ACR. Heuristically, to counterbalance the gravitational influence of
the perturbing planet `b', we need to increase somewhat the mass of the outermost planet
`c'. This assumption is confirmed by numerical integration: orbital configuration with
$m_c$ increased by about $8\%$ shows much less libration amplidute. Moreover, this
adjustment decreases the statistic $\tilde l$ by about $0.1$~m/s (i.e., the scatter of the
data around the RV model becomes slightly less). In the case of the 5/2 resonance,
(solution I$^c$B), the libration amplitude is quite small for the unperturbed ACR solution
already. The evolution of these orbital configurations appears perfectly regular and does
not show any instability at the time scale of at least $10^6$~yr.

More detailed analysis shows a great diversity of dynamical behaviour in the vicinity of
the ACR solutions. The amplitude and character of the librations can be different. For
certain orbital configurations inside the 2/1 MMR, the system may switch (from time to
time) between alternating asymmetric ACRs with $\omega_c-\omega_d\approx -60^\circ$ and
$\omega_c-\omega_d\approx +60^\circ$. Librations surrounding simultaneously the pair of
stable asymmetric stationary solutions and unstable symmetric aligned ACR (see the
left-middle panel in Fig.~\ref{fig_plan}) are also possible.

\section{Testing the existence of extra planets}
\label{sec_extra}
When the best fitting orbital structure of the planetary system appears unstable, we can
suspect that more planets orbit the star. It was the case for the system of $\mu$~Ara: the
determination of realistic orbits of $\mu$~Ara~b and~c represented an essential difficulty
\citep[e.g.,][]{Gozd05} until the discovery of the planet $\mu$~Ara~e placed all in the
places \citep{Pepe07,Gozd07}. To test the hypothesis of an extra planet orbiting HD37124,
we use the likelihood ratio--based periodogram described in the paper \citep{Baluev08b}.
This periodogram represent a generalization of the usual \citet{Lomb76}-\citet{Scargle82}
periodogram (as well as of the data-compensated discrete Fourier transform periodogram by
\citet{FerrazMello81}) and incorporates a built-in estimation of the RV jitter.
\begin{figure}
\begin{tabular}{@{}c@{}c@{}}
  \includegraphics[width=0.51\linewidth]{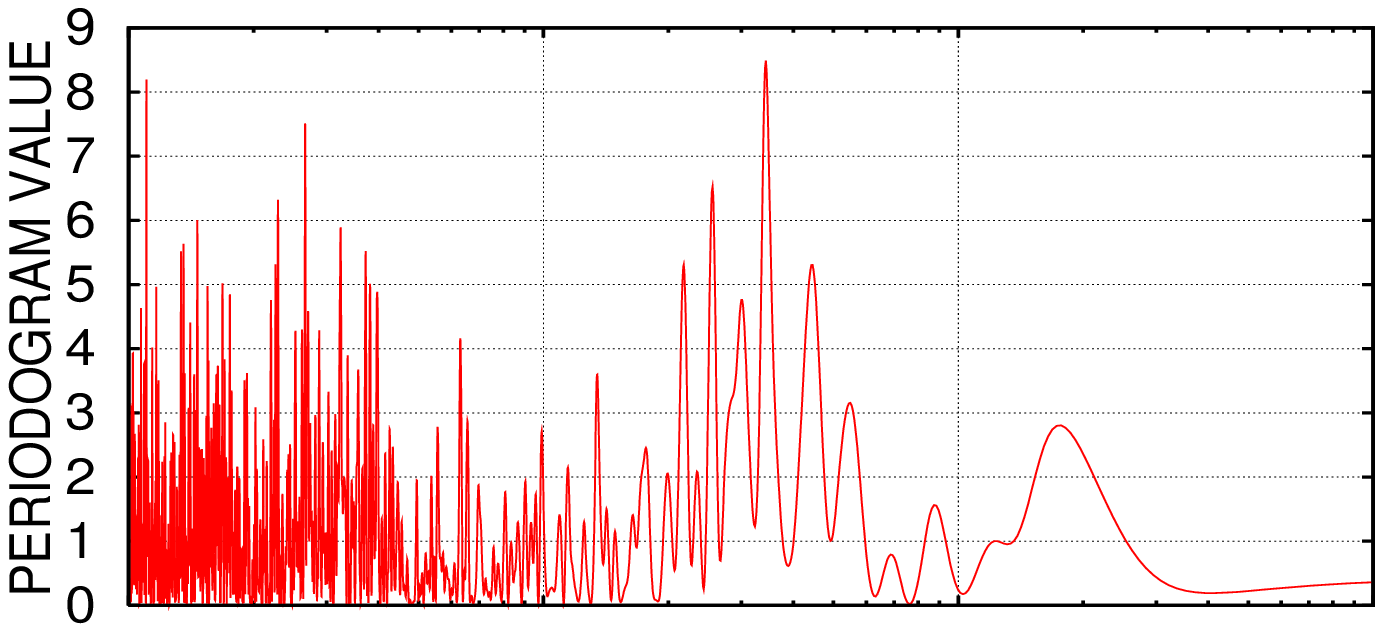} &
  \includegraphics[width=0.49\linewidth]{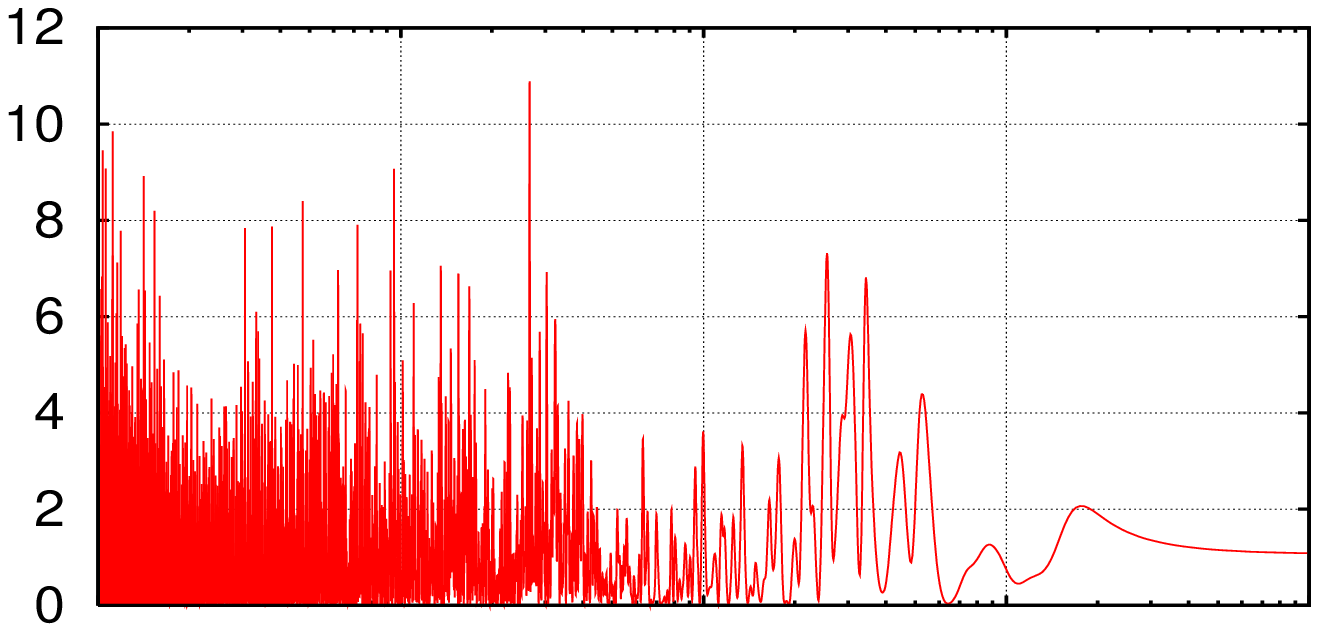} \\
  \includegraphics[width=0.51\linewidth]{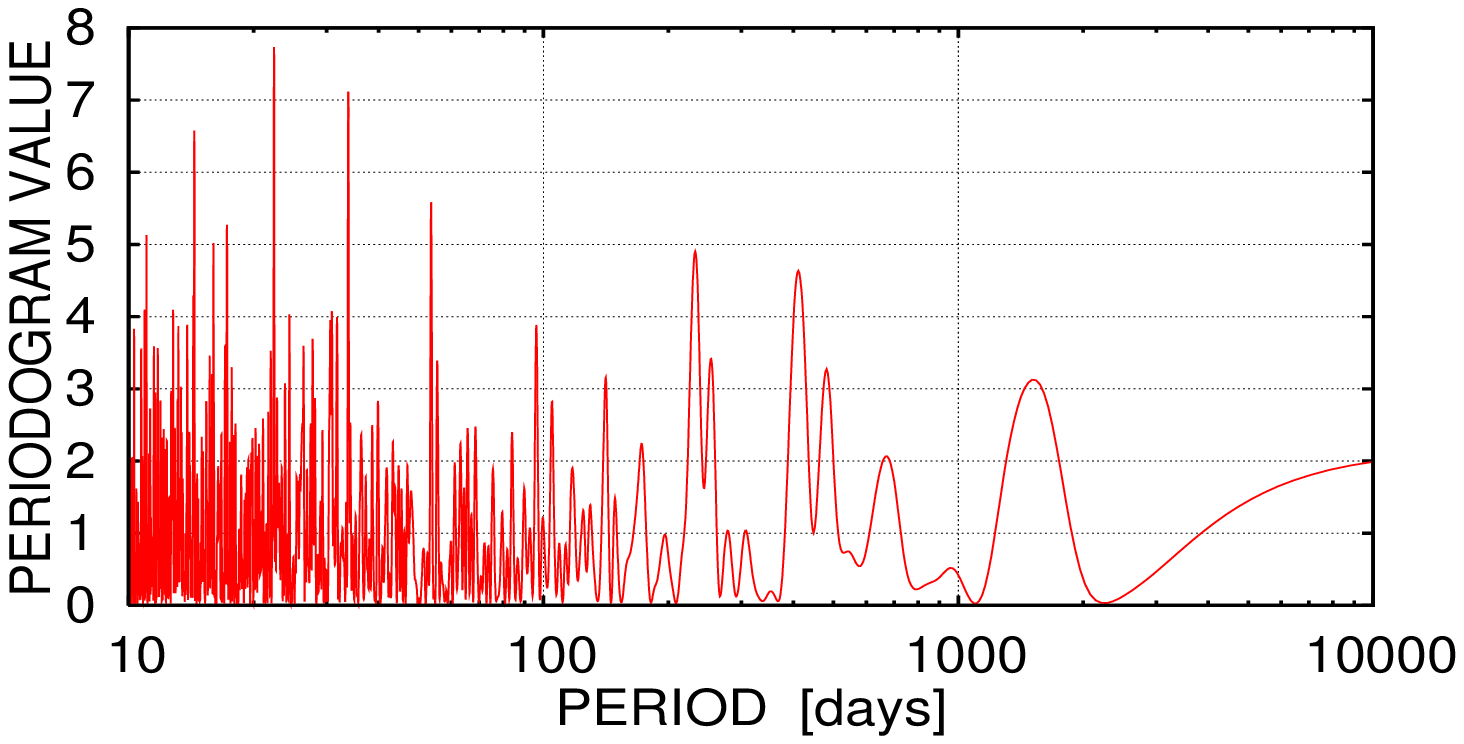} &
  \includegraphics[width=0.49\linewidth]{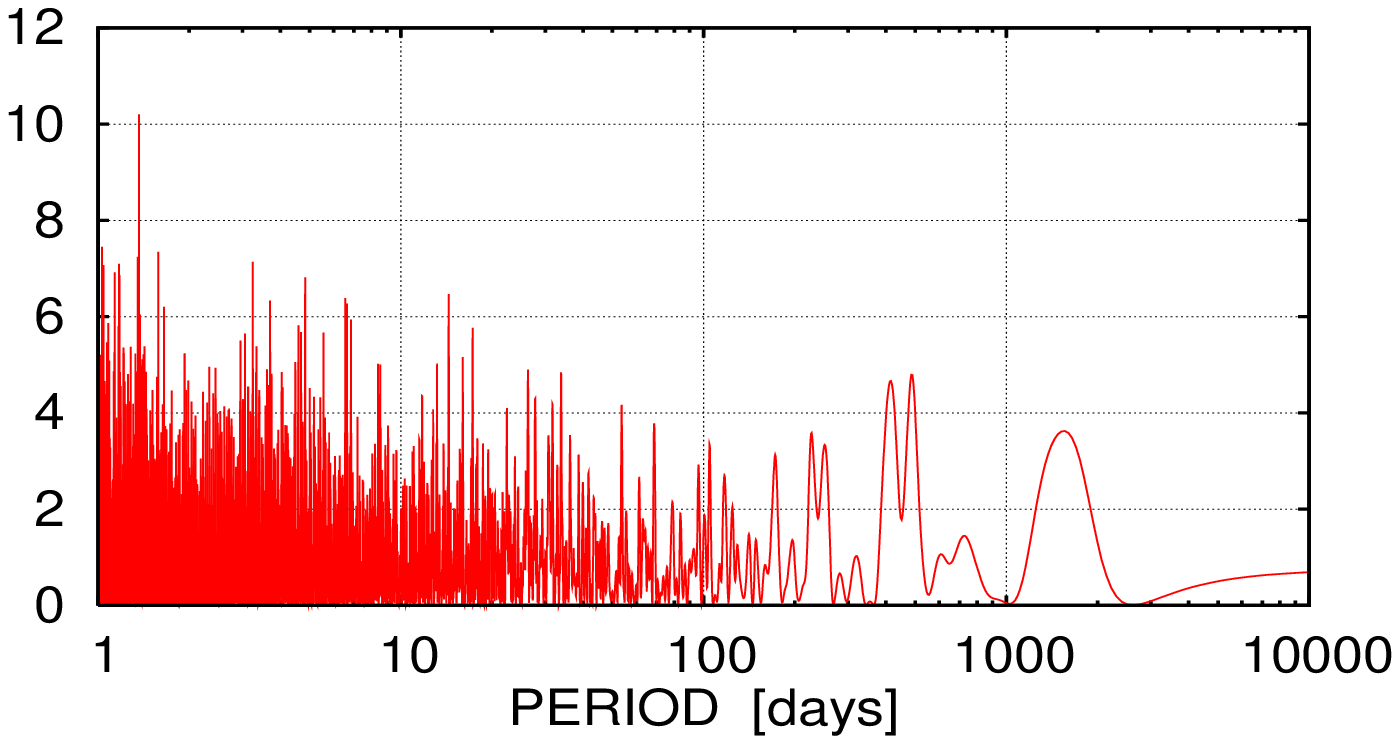} \\
\end{tabular}
\caption{The likelihood ratio periodograms of the RV residuals for different three-planet
orbital models of the system of HD37124. Every value of these periodograms represents the
modified likelihood ratio statistic \citep[see][]{Baluev08b}, calculated for the base
model of the residuals (free constant velocity offsets + free common linear drift) and for
the alternative model incorporating also a sinusoidal variation with free amplitude and
phase. The panels in the left column represent the periodograms constructed from the
residuals of full RV dataset. They were constructed in the range of periods starting from
$10$~days, since the errors in the dates of ELODIE measurements do not allow to fit
accurately more short periods. The graph in the right columns show the periodograms
constructed for the Keck residuals only. The top pair of panels is for the ACR solution in
the 2/1 MMR (RV model~I). The bottom pair of panels is for the ACR solution in the 5/2 MMR
(also RV model~I). For the case of graphs to the left, the normalized frequency bandwidth
$W\approx 350$, and for the graphs to the right $W\approx 3400$.}
\label{fig_period}
\end{figure}

The graphs of such periodograms of RV residuals for several orbital solutions and
involving different datasets are shown in Fig.~\ref{fig_period}. We can see that none of
the peaks rises clearly beyond the apparent noise level. The formal significances of a
periodogram peak can be assessed using the analytic expression for the associated false
alarm probabilities (i.e., the probability to claim that the peak is statistically
significant when actually it is a result of noise fluctuations) from the paper
\citep{Baluev08a}:
\begin{equation}
(\mathrm{false\ alarm\ probability}) \approx W e^{-z} \sqrt z,
\end{equation}
where $z$ is the height the maximum periodogram peak, and $W = f_{\rm max} T_{\rm eff}$ is
the normalized frequency bandwidth, with $f_{\rm max}$ being the maximum frequency being
scanned and $T_{\rm eff}$ being the effective time series span (which is usually close to
the actual time base). The applicability of this formula to the likelihood ratio
periodograms was also discussed in \citep{Baluev08b}. None of the peaks on the
periodograms in Fig.~\ref{fig_period} possesses the false alarm probability estimation
less than $\sim 20\%$. This means that no extra detectable periodicity is present in the
RV data being used in our work. The only suspiciuous peak is shown by the periodogram of
the residuals to the solution I$^c$A (the upper-right panel in Fig.~\ref{fig_period}).
This peak is close to the period of $26.6$~days and possesses the false alarm probability
of $\sim 2\%$, if the latter is calculated for the period range $P\geq 10$~days (instead
of $P\geq 1$~days). However, this peak is not present in other periodograms and thus seems
to be a noise artefact exacerbated by inaccuracies of the model of the RV curve for this
orbital solution and by aliasing (caused by uneven spacings of the RV data), rather than
to reflect an RV oscillation induced by an extra hypothetical planet.

There is also possibility that the putative additional planets are also trapped in a MMR
with one of the two outer planets. For example, the RV oscillation of the hypothetical
fourth planet having orbital period of $P_d/2 \approx 450$~days or $P_c/3 \approx
650$~days would be extremely difficult to extract from the synthetic RV curve. Such RV
oscillation could be almost equally treated as a Fourier overtone associated with
Keplerian oscillations induced by the outer planets (thus resulting in some change of
their best-fitting eccentricities). Since the currently available RV data for HD37124 do
not allow reliable determination of the orbits in the system even for the three-planet
configuration, the more complicated (and worse determined) four-planet configurations were
not considered here. It is worth emphasizing that there is no necessity to call for
four-planet configurations of HD37124. All present difficulties connected with this system
can be explained from the positions of data analysis, that is by the lack of the RV data,
which is not sufficient to constrain reliably the non-linear triple-planet RV model having
large number of degrees of freedom. For example, it was shown by \citet{Baluev08-IAUS249}
that the unrealistically large formal estimations of the eccentricities can be interpreted
as a result of their statistical (systematic) biasing.

\section{Conclusions}
\label{sec_conc}
In the paper, the full set of high-precision RV data available for the planetary system of
HD37124 is analysed. The analysis involves different RV models and accounts for the
requirement of the dynamical stability of the planetary system. The most likely orbital
configurations of the system, found in the paper, are split in four classes:
\begin{enumerate}
\item Two outer planets `c',`d' are captured in the 2/1 mean-motion resonance and move on
orbits with low or moderate eccentricities ($e\lesssim 0.15$). The planets are far from an
apsidal corotation resonance, but the whole system is still stable due to relatively low
eccentricities.
\item The planets `c',`d' are in the 2/1 mean-motion resonance and move on significantly
elliptic orbits with moderate or high eccentricities. The system is stable, because the
planets are locked in (or librate around) an asymmetric apsidal corotation.
\item The planets `c',`d' are in the 5/2 mean-motion resonance. They move on elliptic
orbits with high or moderate eccentricities. The planetary orbits are intersecting (or
close to an intersection), but the system is stable, because the planets are locked in (or
librate around) a symmetric antialigned apsidal corotation. This branch of solutions was
also groped by \citet{Gozd06} basing on the Keck data only.
\item The planets `c',`d' are not necessarily trapped in a mean-motion resonance. They
move on orbits with relatively low eccentricities, which make the whole system stable.
However, these solutions show larger scatter of the RV residuals and thus are less likely
then solutions from other branches. When the eccentricies are fixed at small values, the
ratio of the best fitting orbital periods of the two outer planets appears within a few
per cent of the 2/1 MMR. The values of the period ratio $P_c/P_d$ exceeding $2.3-2.5$ are
unlikely. However, the configurations with $P_c/P_d<2.3$ are quite possible, bearing in
mind the example of the planets b and e in the system of $\mu$~Ara, which are close to,
but likely not trapped in, the 2/1 MMR with $P_b/P_e\approx 2.1$ \citep{Gozd07,Short08}.
\end{enumerate}
The branches (2) and (3) are of a special interest, because we do not know examples of
planetary systems with similar orbital configurations. The possibility of such orbital
configurations makes the algorithms of optimal scheduling of RV observations
\citep{Baluev08d,Ford08} extremely tempting to use for this star.

It was also shown that the RV residuals do not contain any detectable extra periodicity
which could provide clear evidence for a fourth planet in the system. However, for the
solutions close to the 2/1 MMR, there are evidences for an extra variation in the RV
residuals. This variation can be explained by putative annual errors in the Keck data or,
alternatively, by extra linear RV drift which could be induced by a distant unseen
companion of the star.

\begin{acknowledgements}
This work was supported by the Russian Foundation for Basic Research (Grant 06-02-16795)
and by the President Grant NSh-1323.2008.2 for the state support of leading scientific
schools. I am grateful to Profs. K.V.~Kholshevnikov and V.V.~Orlov for useful comments and
correction. I would like to thank the referees, C.~Beaug\'e and the anonymous one, for
careful reading of the manuscript and useful suggestions which helped to improve it.
\end{acknowledgements}



\appendix
\section{Obtaining ACR fits}
\label{sec_fitACR}
Here we describe the procedure of obtaining the ACR fits in more details. The algorithm
that we are about to describe may be useful also for other planetary systems. But before
we proceed, we have to choose some coordinate system. Actually, in the case of HD37124, it
may be checked that the offsets of resulting orbital parameters, referenced in different
coordinate systems, would be quite negligible in the sense of the RV fit quality. However,
the type of the coordinate system should be stated for the purposes of long-term
integrations of the planetary system: the coordinate system used in the integration should
match the given orbital elements. The actual choice of the Jacobi coordinates was
motivated here by the fact noted in \citep{LissRiv01,LeePeale03}, that it is Jacobi
coordinate system in which the osculating orbital elements are mostly close to those
obtained using the kinematic (Keplerian) RV model, especially when the system contains
hierarchical planet pairs (like pairs of planets b--c and b--d in the case of HD37124).
This property of the Jacobi coordinates may be useful, for instance, during a transition
from multi-Keplerian to N-body model of the RV, that was (and probably will be) needed for
some resonant planetary systems after accumulating a sufficiently long observation time
span.

In the Jacobi coordinates, the Hamiltonian of a system with $N$ planets looks like
\begin{equation}
H = \sum_{i=1}^N \left( \frac{{\vec p'_i}^2}{2 m'_i} - \frac{G m_0 m_i}{r_i} \right) -
\sum_{1\leq i<j \leq N} \frac{G m_i m_j}{r_{ij}},
\label{Ham-Jac}
\end{equation}
where $G$ is the gravitational constant, $m_i$ are the planetary masses, $r_i$ are the
astrocentric distances of the planets, $r_{ij}$ are the distances between the planets,
$\vec p'_i$ are the Jacobi momenta and $m'_i = m_i (\sum_{j=0}^{i-1} m_j) /
(\sum_{j=0}^i m_j)$ are the Jacobi masses. Now we need to split~(\ref{Ham-Jac}) in the
Keplerian part that we consider as unperturbed one and in the part that we consider as
perturbational function. This may be done non-uniquely. We adopt the splitting $H =
\sum_{i=1}^N H_{\mathrm{Kep},i} - \sum_{i=1}^N R_i$, where
\begin{equation}
 H_{\mathrm{Kep},i} = \left( \frac{{\vec p'_i}^2}{2 m'_i} - \frac{G M_{i-1}}{r'_i} m_i
 \right),\qquad
  R_i = G m_i \left( \sum_{j=i+1}^N \frac{m_j}{r_{ij}} + \frac{m_0}{r_i} -
  \frac{M_{i-1}}{r'_i} \right).
\end{equation}
Here $r'_i$ is the length of the Jacobi radius-vector for the $i^{\rm th}$ planet,
$M_{i-1}=\sum_{j=0}^{i-1} m_j$ is the sum of the star mass and of the masses of all
planets being interior with respect to the given planet. Therefore, $H_{\mathrm{Kep},i}$
is chosen so that the corresponding unperturbed Keplerian orbit is referenced to a
fictitious central body having mass $M_{i-1}$. We will use only the second-order
approximation of the Hamiltonian. In this approximation, we represent $R_i =
\sum_{j=i+1}^N R_{ij} + \mathcal O(m^3)$, where
\begin{equation}
R_{ij} = G m_i m_j \left( \frac{1}{|\vec r'_i - \vec r'_j|} - \frac{\vec r'_i
\cdot \vec r'_j}{{r'_j}^3} - \frac{1}{r'_j}
\right).
\end{equation}
Here $\vec r'_i$ are the Jacobi position vectors. This approximation for the case of
two planets may be found in \citep[\S~4.2]{Mush}, and the extension to $N\geq 3$ planets
is straightforward.

Let us define the following function of six input parameters:
\begin{equation}
\mathcal R(\alpha, e_{\rm inn}, e_{\rm out}, \lambda_{\rm inn}, \lambda_{\rm out},
\Delta\omega) = \frac{1}{|\vec r_{\rm inn} - \vec r_{\rm out}|} - \frac{\vec
r_{\rm inn}\cdot \vec r_{\rm out}}{r_{\rm out}^3} - \frac{1}{r_{\rm out}}.
\label{funcR}
\end{equation}
Here, the vectors $\vec r_{\rm inn}$ and $\vec r_{\rm out}$ describe positions of
abstract `inner' and `outer' planets moving on Keplerian orbits. The Keplerian orbital
elements marked by the subscript `$\rm inn$' refer to the orbital elements of the inner
planet, and quantitties marked by the subscript `$\rm out$' correspond to the outer
planet. The semi-major axis of the inner abstract planet is set to unit, and the
semi-major axis of the abstract outer planet is equal to $\alpha$. The parameter
$\Delta\omega = \omega_{\rm out}-\omega_{\rm inn}$ represents the difference between the
longitudes of periastra. Then the obvious identity $R_{ij} = \frac{G m_i m_j}{a_i}
\mathcal R(a_j/a_i,e_i,e_j,\lambda_i,\lambda_j,\omega_j-\omega_i)$ holds true.

As usually, the Keplerian parts of the Hamiltonian depend only on the Delaunay actions
$L_i$ (and, as on parameters, on the masses of the planets) and remain unchanged by any
averaging. Let us now restrict our attention to the motion of the (possibly) resonant
planets HD37124 c and d, neglecting the influence of the innermost planet b. The
perturbational function describing the interaction of the planets c and d is given by
$\frac{G m_c m_d}{a_d}\mathcal R(\alpha=a_c/a_d, e_d,e_c, \lambda_d,\lambda_c,
\Delta\omega=\omega_c-\omega_d)$. The corresponding averaged perturbational function is
then given by $\frac{G m_c m_d}{a_d} \langle \mathcal R\rangle (\alpha,e_d,e_c,s_d,s_c)$,
where the averaged function $\langle \mathcal R \rangle$ depends on only five input
parameters $\alpha, e_{\rm inn}, e_{\rm out}, s_{\rm inn}, s_{\rm out}$:
\begin{eqnarray}
\langle \mathcal R\rangle(\alpha,e_{\rm inn}, e_{\rm out}, s_{\rm inn}, s_{\rm out}) =
\int_0^{2\pi} \mathcal R\left(\alpha, e_{\rm inn}, e_{\rm out}, s_{\rm inn}+p\theta,
s_{\rm out}+q\theta, s_{\rm inn}-s_{\rm out} \right) \frac{d\theta}{2\pi}.
\label{avrR-1}
\end{eqnarray}
Note that the last term in the expression~(\ref{funcR}) is averaged to $1/\alpha$. This
term reflects only the fact that the osculating Keplerian orbits are refered to different
fictitious central masses. In fact, we need to average only the resting classical
expression of the perturbational function:
\begin{equation}
\langle\mathcal R \rangle = \left\langle \frac{1}{|\vec r_{\rm inn} - \vec r_{\rm
out}|} - \frac{\vec r_{\rm inn}\cdot \vec r_{\rm out}}{r_{\rm out}^3} \right\rangle
- \frac{1}{\alpha}.
\label{avrR-2}
\end{equation}
We perform the averaging~(\ref{avrR-1},\ref{avrR-2}) by means of numerical integration
tools, as it was proposed by \citet{Michtchenko06}. This way of numerical averaging of the
Hamiltonian is very easy to implement and simultaneously is quite rapid and precise. In
the same way, we can calculate various derivatives of the averaged function $\langle
\mathcal R\rangle$, that we will need below. For example,
\begin{equation}
\alpha \frac{\partial\langle R \rangle}{\partial\alpha} = \left\langle \frac{\vec
r_{\rm inn}\cdot \vec r_{\rm out} - r_{\rm out}^2}{|\vec r_{\rm inn} - \vec
r_{\rm out}|^3} + 2\, \frac{\vec r_{\rm inn}\cdot \vec r_{\rm out}}{r_{\rm out}^3}
\right\rangle +
\frac{1}{\alpha},
\end{equation}
where the averaging can be again performed numerically.

In the next step, we need to build in our RV fitting algorithm the four equality bounds
$\partial\langle H_{cd} \rangle/\partial(I_c,I_d) = 0$ and $\partial\langle H_{cd}
\rangle/\partial(s_c,s_d) = \partial\langle R_{cd} \rangle/\partial(s_c,s_d) = 0$, which
determine the location of the ACR. The second pair of equations reflects the fact that
$s_d$ and $s_c$ should correspond to the extremum value of $\langle \mathcal R
\rangle(\alpha=a_c/a_d,e_d,e_c,s_d,s_c)$ (considering the values of $e_{cd}$ and $\alpha$
fixed). This condition can be used to construct the angles $s_{c,d}$ as functions of the
eccentricities $e_{c,d}$: $s_d = s_{\rm inn}^*(\alpha,e_d,e_c)$ and $s_c = s_{\rm
out}^*(\alpha,e_d,e_c)$, where the functions $s_{\rm inn}^*(\alpha,e_{\rm inn},e_{\rm
out})$ and $s_{\rm out}^*(\alpha,e_{\rm inn},e_{\rm out})$ provide an extremum to
$\langle\mathcal R \rangle(\alpha,e_{\rm inn},e_{\rm out},s_{\rm inn},s_{\rm out})$ given
fixed $\alpha,e_{\rm inn},e_{\rm out}$. When the ACR is symmetric, the functions $s_{\rm
inn}^*$ and $s_{\rm out}^*$ can be found easily (actually, they appear to be constant in
this case). However, for an asymmetric ACR, we have to locate the values of $s_c$ and
$s_d$ numerically. Eventually, this numerical procedure is still sufficiently rapid and
can be implemented as a `black-box' subroutine, which returns the values of $s_{\rm
inn}^*$ and $s_{\rm out}^*$ for input values of $e_{\rm inn},e_{\rm out}$, and $\alpha$.
Note that since we consider here only MMR solutions, $\alpha \approx \alpha_0
\equiv (p/q)^{2/3}$ with a error of $\mathcal O(m_{c,d}/M_\star)$. When calculating the
resonant angles, we can quite neglect such errors and put simply $\alpha = \alpha_0$. Such
errors in resonant angles will not produce significant changes in the dynamics of the
planetary system. The resulting values of $s_c$ and $s_d$ and the
definitions~(\ref{res-ang}) can be used in the work of the fitting algorithm to express
$\omega_c$ and $\omega_d$ via the other free parameters: $\lambda_c,\lambda_d,e_c,e_d$.

Another pair of constraints, $\partial\langle H_{cd}\rangle/\partial(I_c,I_d)=0$, can be
transformed to a more convenient (and equivalent) form, involving partial derivatives over
the Delaunay actions $L_{c,d}$ and $G_{c,d} = L_{c,d}\eta_{c,d}$ (here $\eta_{c,d}^2 =
1-e_{c,d}^2$). The first transformed equation, $\partial \langle H_{cd}\rangle/\partial
G_c = \partial \langle H_{cd}\rangle/\partial G_d$, reflects the coincidence of the
secular drifts of the orbital periastra. After neglecting insignificant errors of the
order of the planetary masses, this equation can be simplified to
\begin{equation}
\frac{\tilde K_d}{\tilde K_c} = \alpha_0 \rho(\alpha_0,e_d,e_c) \quad {\rm with}\quad
\rho(\alpha,e_{\rm inn},e_{\rm out}) = \frac{\partial\langle \mathcal
R\rangle/\partial\eta_{\rm inn}}{\partial\langle
\mathcal R\rangle/\partial\eta_{\rm out}} \Bigg|_{s_{\rm inn} = s_{\rm inn}^*(e_{\rm
inn},e_{\rm out}) \atop s_{\rm out} = s_{\rm out}^*(e_{\rm inn},e_{\rm out})},
\label{KcKd}
\end{equation}
where we note that $\frac{\partial}{\partial\eta} = - \frac{\eta}{e}
\frac{\partial}{\partial e}$. The equality~(\ref{KcKd}) can be used to express the
RV semi-amplitude of one of the resonat planets via the RV semi-amplitude of another one
and via the orbital eccentricities. The second transformed equation, $p\, \partial \langle
H_{cd}\rangle/\partial L_c = q\, \partial \langle H_{cd}\rangle/\partial L_d$, reflects
the vanishing of the secular drift of the critical angle $p l_c - q l_d$ (where $l$ are
the planetary mean anomalies). This should provide the long-term constancy of the
planetary conjunction positions. After some simplifications, this equation may be
rewritten as
\begin{eqnarray}
\frac{P_d}{P_c} \frac{p}{q} = 1 - \frac{m_c}{M_\star} \nu(\alpha_0,e_d,e_c)\qquad {\rm
with}\nonumber\\
\nu(\alpha,e_{\rm inn},e_{\rm out}) =
\left[ \left(\frac{p}{q}\eta_{\rm out} - \eta_{\rm inn} \right) \frac{\partial\langle\mathcal
R\rangle}{\partial \eta_{\rm inn}} - 2 \left( \alpha\frac{\partial \langle\mathcal
R\rangle}{\partial\alpha} \left(1+\frac{p}{q}\rho\right) +
\langle\mathcal R\rangle \right) \right]\Bigg|_{s_{\rm inn} = s_{\rm inn}^*(e_{\rm
inn},e_{\rm out}) \atop s_{\rm out} = s_{\rm out}^*(e_{\rm inn},e_{\rm out})}.
\label{PcPd}
\end{eqnarray}
This equation introduces a error of the second order only, that is $\mathcal
O\left((m/M_\star)^2\right)$. Actually, specifically to the system of HD37124, the small
$\mathcal O(m/M_\star)$ deviation of the period ratio from the exact resonance does not
affect significantly the quality of the RV fit (moreover, the values of $\nu$ for the ACR
fits from Table~\ref{tab_apscorfits} appeared less than $0.1$). Almost the same value of
the RV r.m.s. could be obtained for the simplified equation $P_c = \frac{p}{q} P_d$.
However, it is this $\mathcal O(m/M_\star)$ period ratio deviation that determines the
secular drift of the critical angle $p l_c - q l_d$ and the ACR state of the system.
Therefore, in general case it is necessary to take this deviation into account when
constructing the ACR fits. The equation~(\ref{PcPd}) can be used to express the osculating
orbital period $P_c$ or $P_d$ via the resting free variables. Note that the deviation of
the ratio of the osculating semi-major axes can be written down (to within the first
order) as
\begin{equation}
\frac{\alpha}{\alpha_0} = 1 + \frac{m_c}{M_\star} \frac{2\nu(\alpha_0,e_d,e_c)+1}{3}.
\end{equation}

Therefore, for any given values of the parameters $e_c,e_d,\lambda_c,\lambda_d,\tilde K_d,
P_d$ we can obtain the ACR values of the parameters $\omega_c,\omega_d, K_c$ with a error
of $\mathcal O(m_{c,d}/M_\star)$ and the ACR value of $P_c$ with a error of $\mathcal
O\left((m_{c,d}/M_\star)^2\right)$. This allows us to obtain the best-fitting orbital
solution, which is sufficiently close to an ACR state.

\end{document}